\documentclass[reprint,superscriptaddress,twocolumn,10pt]{revtex4-1}
\usepackage{bbm}
\usepackage{physics}
\usepackage{amsmath}
\usepackage{epsfig}
\usepackage{subfigure,mathrsfs}
\usepackage{array}
\usepackage{amssymb}
\usepackage{braket}
\usepackage{float}
\usepackage{lmodern,amssymb}
\usepackage{physics}
\usepackage[dvipsnames]{xcolor}
\usepackage{lipsum}

\newcommand{\beq}[0]{\begin{equation}}
\newcommand{\eeq}[0]{\end{equation}}

\def\be{\begin{equation}}
\def\ee{\end{equation}}
\def\bea{\begin{eqnarray}}
\def\eea{\end{eqnarray}}
\newcommand{\ba}{\begin{eqnarray}}
\newcommand{\ea}{\end{eqnarray}}
\usepackage{hyperref}

\usepackage[bb=boondox]{mathalfa}

\begin{document}
\title{Top-Ranked Cycle Flux Network Analysis of Molecular Photocells}
\author{Nikhil Gupt}
\affiliation{Indian Institute of Technology Kanpur, 
	Kanpur, Uttar Pradesh 208016, India}
\author{Shuvadip Ghosh}
\affiliation{Indian Institute of Technology Kanpur, 
	Kanpur, Uttar Pradesh 208016, India}
\author{Arnab Ghosh}
\thanks{arnab@iitk.ac.in}
\affiliation{Indian Institute of Technology Kanpur, 
	Kanpur, Uttar Pradesh 208016, India}

\begin{abstract}
We introduce a top-ranked cycle flux ranking scheme of network analysis to assess the performance of molecular junction solar cells. By mapping the Lindblad master equation to the quantum-transition network, we propose a microscopic Hamiltonian description underpinning the rate equations commonly used to characterize molecular photocells. Our approach elucidates the paramount significance of edge flux and unveils two pertinent electron transfer pathways that play equally important roles in robust photocurrent generation. Furthermore, we demonstrate that non-radiative loss processes impede the maximum power efficiency of photocells, which may otherwise be above the Curzon-Ahlborn limit. These findings shed light on the intricate functionalities that govern molecular photovoltaics and offer a comprehensive approach to address them in a systematic way.

\end{abstract}

\maketitle

\section{Introduction}
\par Molecular junctions consisting of a single donor and acceptor, placed between two external leads is an active field of research~\cite{nitzan2013chemical,bittner2010quantum,aradhya2013single,su2016chemical} that combines the fundamental aspects of quantum transport~\cite{scully2010quantum,scully2011quantum,michael2018perspective} and their possible practical implementations~\cite{chen2005book,pop2010energy,killoran2015enhanching,arp2016natural}. One such application of molecular junctions, among many others, is in photovoltaic (PV) cells, where solar energy of incident photons is converted into electric power~\cite{deibel2010polymer,nicholson2010organic,prashant2013quantum}. The interplay between the heat current mediated by the temperature difference of the solar radiation and the PV cell at ambient temperature, together with the charge current arising due to the bias voltage across the electrodes, drive the system towards strong non-equilibrium steady states ~\cite{altaner2012network,lena2012vibrational,agarwalla2015fullcounting,ajisaka2015themolecular}. A great deal is therefore focused on theoretical modeling to optimize the performance of PV cells~\cite{creatore2013efficient,fruchtman2016photocell} and to explore the underlying transport mechanisms that could facilitate the realization of sophisticated on-chip complex quantum thermal devices~\cite{giazotto2012the,joulain2016quantum,whitney2018quantum,fong2019phonon,pekola2021colloquium}.

Though, the master equation in Lindblad form is a conventional theoretical technique for the description of non-equilibrium open quantum systems~\cite{breuer2002book}, particularly in quantum optics~\cite{agarwal2013book} and quantum thermodynamics communities~\cite{sinha2013fluctuation,nikhil2021statistical,gelbwaser2015thermodynamics,kosloff2013quantum,deffner2019quantumthermo,dong2021onthethermodynamics,gupt2022PRE,shuvadip2022univarsal,samarth2023introduction} where spins and atomic degrees are involved, its extension towards molecular systems is relatively a novel area of exploration ~\cite{aviram1974molecular,harbola2006quantum,galperin2007molecular,wang2020combining,subhajit2020environment,oz2023electron}. As an alternative, in recent times, a state space representation in the framework of network theory by Nitzan et. al.~\cite{einax2011heterojunction,einax2013multiple,einax2014network,einax2016maximum,craven2018electron} has become a popular method to study the non-equilibrium charge transport behavior of molecular photovoltaics~\cite{einax2011heterojunction,einax2013multiple,einax2014network,einax2016maximum} and thermoelectric devices~\cite{whitney2018quantum,craven2018electron,mazza2014thermoelectric}. In this context, the latest finding by Wang et al.~\cite{wang2022cycleflux}, based on the works of Zurek~\cite{zurek1981pointer}, and Cao~\cite{jianlan2013higher}, is worth mentioning. They have shown that the quantum Lindblad master equation can be cast into a Pauli master equation without loss of any generality. This facilitates the representation of dissipative quantum dynamics as a weighted network with nodes and edges, where nodes (vertices) denote quantum states and edges denote the non-equilibrium transition from one quantum state to another with nonzero flux rates~\cite{chernyak2008pumping,rahav2009directed}. Thus, optimization of the performance of multi-component quantum thermal devices reduces to identifying the major working cycles amongst various possible pathways as is the case with photovoltaics  with a multitude of electron transfer channels. Although the concept of cycle flux was well developed in algebraic graph theory~\cite{hill1975stochastic,kohler1980thefrequency,schnakenberg1976network,tutte2001book,balakrishnan2012book,polettini2014transient,owen2020universal} from the early works of Hill~\cite{hill1975stochastic}, Kohler and Vollmerhaus~\cite{kohler1980thefrequency}, as well as Schnakenberg~\cite{schnakenberg1976network}, the recent developments by Wang et. al. have provided an efficient cycle flux ranking scheme to fully comprehend the intricate functionality of complex quantum systems, with particular emphasis on spin-Seebeck effect within the linear response regime~\cite{wang2022cycleflux} .

In this paper, we extend this idea to molecular systems and establish the equivalence between the dynamical formulation of the Lindblad master equation and the state space representation of molecular photocells pioneered by Nitzan et. al.~\cite{einax2011heterojunction,einax2013multiple,einax2014network,einax2016maximum}. We point out however the state space method is quite effective in computing the steady-state currents in PV cells, it falls short to unravel the underlying working mechanism of the photovoltaic devices. On the contrary, top-ranked cycle flux analysis provides a natural and alternative gateway to capture the underlying features, which is otherwise challenging due to multiple electron transfer pathways in molecular junctions. Thus, our present findings demonstrate that the cycle flux ranking scheme could go beyond its standard applications of near-equilibrium situations and could serve as a potential candidate for decoding the fundamental working principle of complex molecular systems even far from equilibrium scenarios.

The work is organized as follows: In section~\ref{Sec.2}, we introduce the basic model of the PV cell and derive the open quantum dynamics of Lindbladian form which is shown to be equivalent to the Pauli master equations employed by the Nitzan et. al.~\cite{einax2011heterojunction} within a state space formulation. Next, we elaborate on the fundamental principles of the cycle flux ranking scheme in the context of our present model in Sec.~\ref{Sec.III}, and summarize important outcomes and findings of our analysis in Sec.~\ref{Sec-IV}. Finally, we conclude in Sec.~\ref{Sec.V}.

\section{Model and Dynamics}\label{Sec.2}
\begin{figure}
    \centering
    \includegraphics[width=\columnwidth]{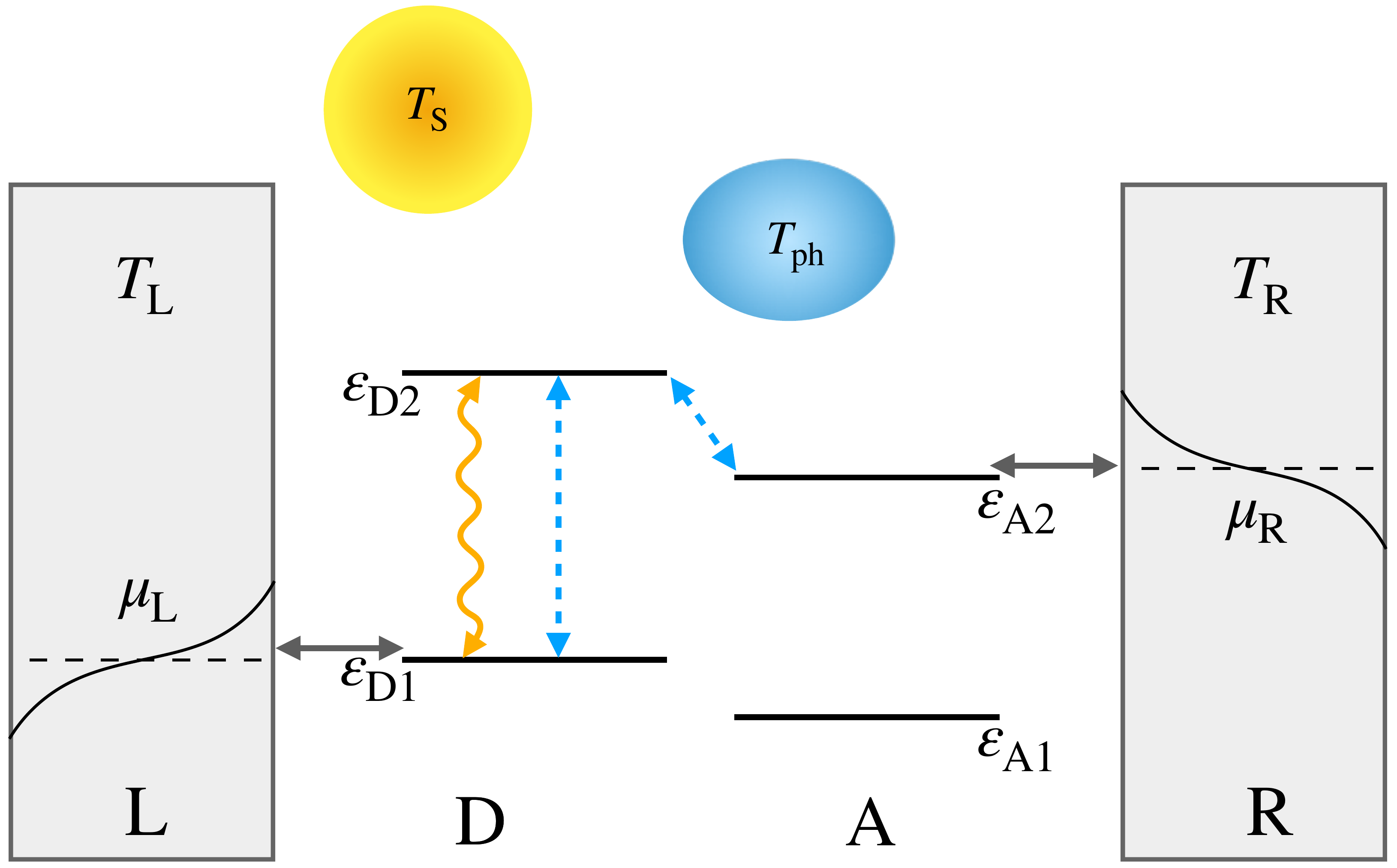}
    \caption{Schematic diagram of a molecular photovoltaic solar cell. The system consists of the donor (D) and the acceptor (A) molecules placed between two free-electron reservoirs. The left reservoir ($\rm L$) is exclusively coupled to  the ground state of the donor and the right reservoir ($\rm R$) is coupled only to  the excited state of the acceptor. Photon-induced transition is indicated by a wiggly line and phonon-induced processes are depicted by broken lines.}
    \label{fig1}
\end{figure}

The basic model of a PV cell is comprised of two ``effective" sites, representing the donor (D) and acceptor (A) molecules placed between two metallic leads (L and R), as depicted in Fig.~\ref{fig1}. We consider each site as a two-state system with the ground ($|n_{\rm s1}\rangle$) and excited ($|n_{\rm s2}\rangle$) states, corresponding to the highest occupied molecular orbital (HOMO) and lowest unoccupied molecular orbital (LUMO) of two molecules having energies $\varepsilon_{\rm s1}$ and $\varepsilon_{\rm s2}$ (${\rm s = D, A}$) respectively. The general form of the electronic part of the Hamiltonian for the donor-acceptor system reads as
\begin{equation}\label{H_S}
    H_{el}= \sum_{\rm s=D,A}\sum_{\rm i=1,2}\varepsilon_{\rm si} n_{\rm si}+ \sum_{\rm s=D,A} U_{\rm s} n_{\rm s1} n_{\rm s2}.  %+ U'n_{AH} n_{AL} (1-n_{DH}) (1-n_{DL}).
\end{equation}
Here, $U_{\rm D}$ $(U_{\rm A})$ stands for the positive Coulomb repulsion energy if two electrons are present in the donor (acceptor) sites and
$n_{\rm si}=c^\dagger_{\rm si} c_{\rm si}$, ${\rm i}= \{1,2\}$ represents the number operator corresponding to the  state $|n_{\rm si}\rangle$, satisfying fermionic anti-commutation relation~\cite{ghosh2012fermionic} $\{c_{\rm si},c^\dagger_{\rm s'j}\}=\delta_{\rm {ss'}}\delta_{\rm {ij}}$. The most important energy scales of the problem are notably the energy gap between the donor levels ($\Delta E = \varepsilon_{\rm D2} - \varepsilon_{\rm D1}$) and the donor-acceptor levels ($\Delta \varepsilon = \varepsilon_{\rm D2} - \varepsilon_{\rm A2}$), while the acceptor's ground state energy has no influence on the overall cell operation as one expects~\cite{einax2011heterojunction,subhajit2020environment}. This can be attributed to the following two facts (i) solar radiation triggers the radiative transition of electrons from the donor ground to the excited state; and (ii) electron transfer between the excited states of the donor to the acceptor could in principle be governed by the cell's vibrational degrees of freedom, such as thermal phonons modes. To make a further realistic assumption about our model in view of the prior literature~\cite{einax2011heterojunction,einax2013multiple,einax2014network,subhajit2020environment}, we restrict the donor to be either in the ground or excited state (i.e., $n_{\rm D1}n_{\rm D2}=0$), whereas the acceptor can be either in the singly or doubly occupied state. As the ground state of the acceptor is always occupied (i.e. $n_{\rm A1}=1$), it can be either in state $|n_{\rm A1} n_{\rm A2}\rangle$=$ |10\rangle$ or in state $|n_{\rm A1} n_{\rm A2}\rangle$=$ |11\rangle$, while, the donor state $|n_{\rm D1} n_{\rm D2}\rangle$ can exist in any one of the three possible configurations $\{ |00\rangle, |10\rangle, |01\rangle \}$. This allows us to directly compare our results with the state space model of Nitzan et. al.~\cite{einax2011heterojunction}, characterized by a single Coulomb repulsion parameter arising out of $U_{\rm{A}}$ in Eq.~\eqref{H_S}.

The Hamiltonian of the free-electron reservoirs~\cite{shuvadip2022univarsal} to which the donor and acceptor molecules are coupled are respectively given by
\begin{equation}
H_{\rm L}=\sum_{l} (\epsilon_{l}-\mu_{\rm L}) d^\dagger_{l} d_{l}, \quad \text{and} \quad H_{\rm R}=\sum_r (\epsilon_r -\mu_{\rm R}) d^\dagger_r d_r, 
\end{equation}
where $d_l(d_r$) and $d^\dagger_l(d^\dagger_r$) are the electron annihilation and creation operators for left (L) and right (R) electrodes and $\mu_\alpha$ is the corresponding chemical potential of $\alpha$-th ($\alpha={\rm L, R}$) reservoirs. In other words, the bias voltage between two metallic electrodes is given by $U=(\mu_{\rm L} -\mu_{\rm R})/|e|$, where $e$ is the charge of the electron. Tunneling Hamiltonian between the electrodes and the molecule is chosen to be of the form of~\cite{lena2012vibrational,shuvadip2022univarsal}
\begin{equation}\label{H_I}
    H_I=\sum_l \hbar g_l(d^\dagger_l c_{\rm D1}+c^\dagger_{\rm D1} d_l) + \sum_r \hbar g_r(d^\dagger_r c_{\rm A2}+c^\dagger_{\rm A2} d_r), 
\end{equation}
where $g_{l(r)}$ is the respective coupling strength. The first term indicates that the left electrode swaps electrons solely with the ground state of the donor which gets triggered by the solar radiation to its excited state. The second term denotes that the right electrode can only exchange electrons with the excited state of the acceptor, as the ground state of the acceptor molecule is already occupied. Finally, electron transfer processes within the system are governed by both photons ($H_{\rm pht}$) and phonon ($H_{\rm phn}$) baths~\cite{subhajit2020environment} with $H_{\rm pht}=\sum_k\epsilon_k a^\dagger_k a_k$, and $H_{\rm phn}=\sum_q \epsilon'_q b^\dagger_q b_q$, where $a^\dagger_k$ ($b^\dagger_q$) and $a_k$ ($b_q$) are bosonic creation and annihilation operators for the $k$-th ($q$-th) bath mode with energy $\epsilon_k$ ($\epsilon'_q$) respectively. The interactions between the photon and phonon baths with the molecule are taken as~\cite{lena2012vibrational,ajisaka2015themolecular,subhajit2020environment}
\begin{equation}\label{photon_H_int}
    H^{\rm pht}_I = \sum_k \hbar g^{\rm D}_{k}(a^\dagger_k c^\dagger_{\rm D1}c_{\rm D2}+c^\dagger_{\rm D2}c_{\rm D1} a_k),
\end{equation}
and
\begin{eqnarray}\label{phonon_H_int}
    H^{\rm phn}_I &=& \sum_q \hbar g^{\rm D}_{q}(b^\dagger_q c^\dagger_{\rm D1}c_{\rm D2}+c^\dagger_{\rm D2}c_{\rm D1} b_q) \nonumber \\
    &+& \sum_q \hbar g^{\rm DA}_{q}(b^\dagger_q c^\dagger_{\rm A2}c_{\rm D2}+c^\dagger_{\rm D2}c_{\rm A2} b_q), 
\end{eqnarray}
respectively, where $g^{\rm D}_k$ and $g^{\rm D}_{q} (g^{\rm DA}_{q})$ are the corresponding coupling constants. From Eqs.~\eqref{photon_H_int} and \eqref{phonon_H_int}, it is clear that the transition between the ground to the excited state at the donor site is governed by both photon (radiative process) and phonon (nonradiative process) modes, while the electron transfer between the donor and the acceptor, is solely driven by vibrational thermal phonon modes at the ambient temperature.

Consequently, the time evolution of the system dynamics in the interaction picture under the Born-Markov approximation is described by the quantum master equation~\cite{breuer2002book,kosloff2013quantum,gupt2022PRE,shuvadip2022univarsal} (Appendix~\ref{Appendix-A})
\begin{eqnarray}\label{Lindblad_ME}
    \frac{d\rho}{dt}=\mathcal{L}_{\rm L}[\rho]+\mathcal{L}_{\rm R}[\rho]+\mathcal{L}^{\rm D}_{\rm pht}[\rho] + \mathcal{L}^{\rm D}_{\rm phn}[\rho]  +\mathcal{L}^{\rm DA}_{\rm phn}[\rho],
\end{eqnarray}
where $\rho$ is the reduced density matrix  of the system and $\mathcal{L}_\nu$ is the Lindblad superoperator describing the effect of dissipation induced by the $\nu$-th thermal bath. Since the Hamiltonian in Eq.~\eqref{H_S} is diagonal in the number state basis for the donor and acceptor molecules, the reduced density matrix $\rho$ of the above Lindblad master equation effectively decouples the diagonal and off-diagonal matrix elements in the eigenbasis of $H_{el}$~\cite{shuvadip2022univarsal}. This, in turn, allows for a closed-form equation of motion for the occupation probabilities or the population of the $\mathbb{i}$-th site as $P_\mathbb{i}=\langle \mathbb{i} | \rho| \mathbb{i} \rangle$, w.r.t the various system eigenstates $\{|\mathbb{i} \rangle \}$. As we mentioned before, the acceptor can only be in two
states $\{|10\rangle, |11\rangle\}$, while the donor can be any of the three
possible configurations $\{|00\rangle, |10\rangle, |01\rangle\}$, so there are six
possible eigenstates for the overall system. We label them as follows: $|\mathbb{0}\rangle=|0010\rangle$, $|\mathbbm{1}\rangle=|1010\rangle$, $|\mathbb{2}\rangle=|0110\rangle$, $|\mathbb{3}\rangle=|0011\rangle$, $|\mathbb{4}\rangle=|1011\rangle$ and $|\mathbb{5}\rangle=|0111\rangle$. The corresponding eigen-energies  $\varepsilon_{\mathbb{j}}$ ($\mathbb{j}=\mathbb{0,1,2, ..,5}$) for the states are given by: $\varepsilon_\mathbb{0}=\varepsilon_{\rm A1}$, $\varepsilon_\mathbb{1}=\varepsilon_{\rm D1}+\varepsilon_{\rm A1}$, $\varepsilon_\mathbb{2}=\varepsilon_{\rm D2}+\varepsilon_{\rm A1}$, $\varepsilon_\mathbb{3}=\varepsilon_{\rm A1}+\varepsilon_{\rm A2}+U_{\rm A}$, $\varepsilon_\mathbb{4}=\varepsilon_{\rm D1}+\varepsilon_{\rm A1}+\varepsilon_{\rm A2}+U_{\rm A}$ and $\varepsilon_\mathbb{5}=\varepsilon_{\rm D2}+\varepsilon_{\rm A1}+\varepsilon_{\rm A2}+U_{\rm A}$ respectively. As a result, the time evolution equations for the population are governed by the kinetic equations (detailed derivation in Appendix~\ref{Appendix-A}) 
\begin{equation}\label{eq_P0}
     \frac{dP_\mathbb{0}}{dt}=(k_{\mathbb{01}}P_\mathbb{1}-k_{\mathbb{10}}P_\mathbb{0})+(k_{\mathbb{03}}P_\mathbb{3}-k_{\mathbb{30}}P_\mathbb{0}),
\end{equation}
\begin{eqnarray}\label{eq_P1}
    \frac{dP_\mathbb{1}}{dt}=(k_{\mathbb{10}}P_\mathbb{0}-k_{\mathbb{01}}P_\mathbb{1})+(k_{\mathbb{12}}P_\mathbb{2}-k_{\mathbb{21}}P_\mathbb{1}) \nonumber\\
    +(k_{\mathbb{14}}P_\mathbb{4}-k_{\mathbb{41}}P_\mathbb{1}),
\end{eqnarray}   
\begin{eqnarray}
    \frac{dP_\mathbb{2}}{dt}=(k_{\mathbb{21}}P_\mathbb{1}-k_{\mathbb{12}}P_\mathbb{2})+(k_{\mathbb{23}}P_\mathbb{3}-k_{\mathbb{32}}P_\mathbb{2}) \nonumber\\
                    +(k_{\mathbb{25}}P_\mathbb{5}-k_{\mathbb{52}}P_\mathbb{2}),
\end{eqnarray}
\begin{eqnarray}
    \frac{dP_\mathbb{3}}{dt}=(k_{\mathbb{30}}P_\mathbb{0}-k_{\mathbb{03}}P_\mathbb{3})+(k_{\mathbb{32}}P_\mathbb{2}-k_{\mathbb{23}}P_\mathbb{3}) \nonumber\\
                        +(k_{\mathbb{34}}P_\mathbb{4}-k_{\mathbb{43}}P_\mathbb{3}),
\end{eqnarray}
\begin{eqnarray}
    \frac{dP_\mathbb{4}}{dt}=(k_{\mathbb{41}}P_\mathbb{1}-k_{\mathbb{14}}P_\mathbb{4})+(k_{\mathbb{43}}P_\mathbb{3}-k_{\mathbb{34}}P_\mathbb{4})\nonumber\\
                    +(k_{\mathbb{45}}P_\mathbb{5}-k_{\mathbb{54}}P_\mathbb{4}),
\end{eqnarray}
\begin{eqnarray}\label{eq_P5}
    \frac{dP_\mathbb{5}}{dt}=(k_{\mathbb{52}}P_\mathbb{2}-k_{\mathbb{25}}P_\mathbb{5})+(k_{\mathbb{54}}P_\mathbb{4}-k_{\mathbb{45}}P_\mathbb{5}).
\end{eqnarray}
This is a ``classical'' looking Pauli master equation where the transition rates involved are quantum mechanical in nature. For instance, $k_{\mathbb{ji}}$ ($k_{ \mathbb{j}\leftarrow \mathbb{i}}$) denotes the rate of transition from quantum state $|\mathbb{i}\rangle$ to $|\mathbb{j}\rangle$ ($\mathbb{i}, \mathbb{j}= \mathbb{0,1,2,3,4,5}$ but $\mathbb{i} \neq \mathbb{j}$), which are given by:
\begin{eqnarray}   
   k_{\mathbb{10}}&=&k_{\mathbb{43}}=\gamma_{\rm L} f(\epsilon_{\rm L}) \label{rate_kL}, \\
    k_{\mathbb{01}}&=&k_{\mathbb{34}}=\gamma_{\rm L} [1-f(\epsilon_{\rm L})],\\
    k_{\mathbb{30}}&=&k_{\mathbb{41}}=k_{\mathbb{52}}=\gamma_{\rm R} f(\epsilon_{\rm R}), \\
    k_{\mathbb{03}}&=&k_{\mathbb{14}}=
    k_{\mathbb{25}}=\gamma_{\rm R} [1-f(\epsilon_{\rm R})],\\
    k_{\mathbb{21}}&=&k_{\mathbb{54}}\equiv k_{\rm r}+k_{\rm nr} \nonumber,\\
          &=&\gamma^{\rm D}_{\rm pht} n(\epsilon_{\rm r})+\gamma^{\rm D}_{\rm phn} n(\epsilon_{\rm nr}),\\
    k_{\mathbb{12}}&=&k_{\mathbb{45}}\equiv \Tilde{k}_{\rm r}+\Tilde{k}_{\rm nr} \nonumber,\\
          &=&\gamma^{\rm D}_{\rm pht} [n(\epsilon_{\rm r})+1]+\gamma^{\rm D}_{\rm phn} [n(\epsilon_{\rm nr})+1],\\
    k_{\mathbb{23}}&=&\gamma^{\rm DA}_{\rm phn} n(\epsilon_{\rm DA}), \\
    k_{\mathbb{32}}&=&\gamma^{\rm DA}_{\rm phn} [n(\epsilon_{\rm DA})+1] \label{rate_kDA}.
\end{eqnarray}
The transition rates consist of two terms: (a) the rate coefficient $\gamma$ depends on the coupling strength through the bath spectral function that characterizes the inverse time scale associated with the corresponding processes; (b) the second term contains the information about the statistical properties of the quantum bath (fermionic and bosonic) through its temperature-dependent auto-correlation functions~\cite{ghosh2012fermionic}. It can be classified into two categories, namely, the absorption and relaxation processes. For a fermionic bath, the  absorption or excitation process is governed by $f(\epsilon_\alpha)$, while the de-excitation process is controlled by $1-f(\epsilon_\alpha)$ factor~\cite{nikhil2021statistical,shuvadip2022univarsal}. The $f(\epsilon_\alpha)$ is the Fermi-Dirac distribution which is given by $f(\epsilon_\alpha)= {1}/(e^{\epsilon_\alpha}+1) \equiv f(\varepsilon_{\mathbb{j}\mathbb{i}},\mu_\alpha,T_\alpha)$, where $\alpha = {\rm L}, {\rm R}$, $\epsilon_\alpha=(\varepsilon_{\mathbb{j}\mathbb{i}}-\mu_\alpha)/k_B T_\alpha$ and $T_\alpha$ is the temperature of the $\alpha$-th reservoir. In the case of bosonic (photon and phonon) bath, the same factors are given by $n(\epsilon_\beta)$ and $1+n(\epsilon_\beta)$~\cite{nikhil2021statistical}, with  $n(\epsilon_\beta)$ as the Bose-Einstein distribution $n(\epsilon_\beta)={1}/(e^{\epsilon_\beta}-1)$, for $\beta={\rm r},{\rm nr},{\rm DA}$, where $\epsilon_{\rm r}=(\varepsilon_{\rm D2}-\varepsilon_{\rm D1})/k_B T_{\rm S}$, $\epsilon_{\rm nr}=(\varepsilon_{\rm D2}-\varepsilon_{\rm D1})/k_B T_{\rm ph}$, $\epsilon_{\rm DA}=(\Delta\varepsilon-U_{\rm A})/k_B T_{\rm ph}$ and $T_{\rm S}$ ($T_{\rm ph}$) stands for the temperature of the photon (phonon) bath respectively.

We emphasize that Eqs.~\eqref{eq_P0}-\eqref{eq_P5}, along with their rate coefficients calculated from the above microscopic picture, are analogous to the phenomenological rate equations considered by Nitzan et. al.~\cite{einax2011heterojunction,einax2013multiple,einax2014network} in their study of molecular photocells. In terms of the occupation probabilities of the individual states, empirical definitions for the electron currents used by Nitzan et. al.~\cite{einax2011heterojunction}, within the state-space model, can be summarized as 
\begin{eqnarray}
J_{\rm R} &=& [k_{\mathbb{30}}P_\mathbb{0}-k_{\mathbb{03}}P_\mathbb{3}]+[k_{\mathbb{41}}P_\mathbb{1}-k_{\mathbb{14}}P_\mathbb{4}] \nonumber\\
    &+&[k_{\mathbb{52}}P_\mathbb{2}-k_{\mathbb{25}}P_\mathbb{5}],\\
    J_{\rm L} &=& [k_{\mathbb{10}}P_0-k_{\mathbb{01}}P_\mathbb{1}]+[k_{\mathbb{43}}P_\mathbb{3}-k_{\mathbb{34}}P_\mathbb{4}],\\   
    J_{\rm S} &=& k_{\rm r}[P_\mathbb{1}+P_\mathbb{4}]-\Tilde{k}_{\rm r}[P_\mathbb{2}+P_\mathbb{5}], \\
    J_{\rm nr}&=&  k_{\rm nr}[P_\mathbb{1}+P_\mathbb{4}]-\Tilde{k}_{\rm nr}[P_\mathbb{2}+P_\mathbb{5}],  \\
    J_{\rm DA}&=& k_{23}P_\mathbb{3} - k_{\mathbb{\mathbb{32}}}P_\mathbb{2}.
\end{eqnarray}
Here $J_{\rm R}$ and $J_{\rm L}$ are the electron currents leaving and entering the molecular system to and from the electrodes, $J_{\rm S}$ and $J_{\rm nr}$ are, respectively, the radiative (photon-induced) and nonradiative (phonon induced losses) electron currents between the ground and excited states of the donor, and $J_{\rm DA}$ is the average current due to transfer of electrons (phonon induced) between donor and acceptor molecules. While the above rate equations can be solved numerically to obtain the steady-state flux-current, $J_{\rm L}=-J_{\rm R}=-J_{\rm {DA}}=J_{\rm S}+J_{\rm {nr}}=J$, it does not reveal much about the actual mechanisms of the transport processes involved. In what follows, we take advantage of the cycle flux analysis scheme of the algebraic graph theory~\cite{hill1975stochastic,kohler1980thefrequency,schnakenberg1976network,tutte2001book,balakrishnan2012book,polettini2014transient,owen2020universal} to gain a deeper insight into the underlying transport channels of the molecular solar cells and explore the effect of non-radiative losses on the thermodynamic efficiency of the photocell.

\section{The Cycle flux Analysis}\label{Sec.III}
\begin{figure}
    \includegraphics[width=\columnwidth]{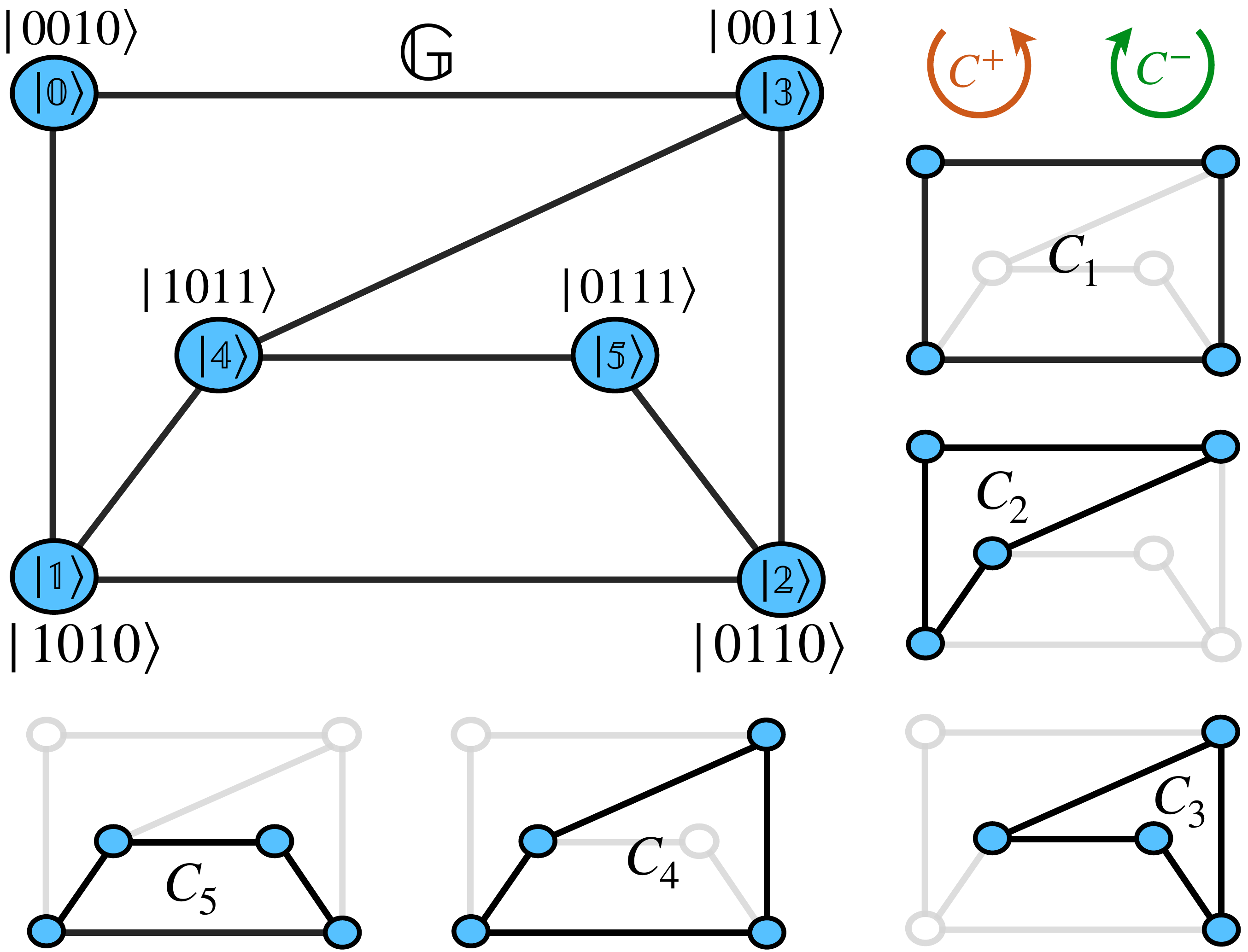}
    \caption{The fundamental basic graph ($\mathbb{G}$) as well as its undirected subcycles, encompasses all possible quantum transport channels (networks) of the photovoltaic solar cell under non-equilibrium conditions.}
    \label{fig2}
\end{figure}

From Eqs.~\eqref{eq_P0}-\eqref{eq_P5}, we may rewrite the time evolution equation of the population distribution $P_\mathbb{i}$ over the quantum mechanical system states in the following compact form
\begin{eqnarray}\label{master-eqn-compact}
    \frac{dP_\mathbb{i}}{dt}=\sum^\mathbb{5}_{\mathbb{j=0}}[k_{\mathbb{ij}}P_\mathbb{j}-k_{\mathbb{ji}}P_\mathbb{i}];\quad \mathbb{j}\ne{\mathbb{i}},
    \label{mastereqn}
\end{eqnarray}
subject to the condition $\sum_\mathbb{i=1}^{\mathbb 5} P_\mathbb{i}=1$. Thus it completely characterizes the overall system where $k_{\mathbb{ji}}$ ($k_{\mathbb{ij}}$) depicts the open quantum system under non-equilibrium conditions with forward (backward) transition rates listed in Eqs.~\eqref{rate_kL}-\eqref{rate_kDA}. With the help of higher-order kinetic expansion, dissipative dynamics in presence of quantum coherence can be even mapped onto kinetic network~\cite{jianlan2013higher}. Therefore, without loss of any generality, and following the Ref.~\cite{wang2022cycleflux}, the above dissipative quantum dynamics can be effectively represented by a network or graph, where nodes or vertices are assigned to each quantum state, and edges or lines relate the allowed transitions between them. In the present case, the basic graph ($\mathbb{G}$) depicted in Fig.~\ref{fig2} provides a useful visualization of the system, with each vertex denoting a quantum state and its associated probability $P_{\mathbb{i}}$, and each edge representing a pair of transition probabilities $k_{\mathbb{ji}}$ and $k_{\mathbb{ij}}$, one for each possible direction. Since the existence of the forward transition rate $k_{\mathbb{ji}}>0$ implies the reverse transition $k_{\mathbb{ij}}>0$ in all practical circumstances, we may assume without loss of any generality that our basic graph $\mathbb{G}$ is connected in the sense that for each pair of states or vertices $(\mathbb{i},\mathbb{j}), \mathbb{i} \neq \mathbb{j}$, there exists \textit{at least one sequence of transitions} or edges that connects them in both directions. If $\mathbb{G}$ is not connected, the physical system represented by $\mathbb{G}$ can be decomposed into non-interacting subsystems, which can be analyzed independently~\cite{schnakenberg1976network}.

Despite the fact that the underlying master equation~\eqref{master-eqn-compact} is linear and the uniqueness of the steady-state solution is guaranteed by the properties of the master equation, finding out the complete analytical solutions of the steady-state populations is a nontrivial task even for the simplest physical conditions~\cite{schnakenberg1976network}. However, the diagrammatic representation of the system in terms of its basic graph offers a highly versatile and effective tool to tackle such complicated problems~\cite{ren2017detectable,annwesha2020stochastic,wang2022cycleflux}. For instance, the steady state solution of $P_\mathbb{i}$ of our basic graph, is defined as~\cite{hill1975stochastic,kohler1980thefrequency,schnakenberg1976network} 
\begin{equation} 
\bar{P}_\mathbb{i}=\Lambda_\mathbb{i}/\Lambda,
\end{equation}
where $\Lambda_{\mathbb{i}}$ is the sum of the weight of the spanning trees rooted on $\mathbb{i}^{th}$ state and $\Lambda$ is the sum of weights of spanning trees rooted on every individual state, i.e., $\Lambda=\sum^{\mathbb{5}}_{\mathbb{i=0}}\Lambda_\mathbb{i}$. The above method was first invented by Kirchhoff in 1847~\cite{kirchhoff1847} in the context of network theory and later on, rigorously formulated by King and Altmann~\cite{king1956schematic} in the context of biochemical reactions. As a result, it is commonly referred to as Kirchhoff's theorem in the theory of network analysis~\cite{kirchhoff1847,king1956schematic,schnakenberg1976network,annwesha2020stochastic}. Here by \textit{spanning tree}, we refer to a covering subgraph of the basic graph that contains all the vertices with the minimum number of edges which is always connected but contains no \textit{circuits} (cyclic sequence of edges or \textit{cycle trajectories}). From Fig.~\ref{fig2}, it is easy to understand that a basic graph generally contains a large number of undirected subcycles, and each subcycle is a combination of a pair of two one-directional \textit{circuits} or cycle trajectories, $C^{+}$ (counterclockwise) and $C^{-}$ (clockwise). The net cycle flux is therefore given by  the difference between the two circuit fluxes such as $J_C=J^+_{C}-J^-_{C}$. The notion of the ``circuit flux" was introduced by Kohler and Vollmerhaus~\cite{kohler1980thefrequency}, and has since been widely employed to characterize a large variety of biological systems~\cite{qian2007phospho,ren2017detectable,barato2015thermodynamic,annwesha2020stochastic,king1956schematic,giebink2011thermodynamic,rutten2009reaching}. In essence, it captures the frequency of circuit completion along a specific cycle trajectory ($C^{+}$ or $C^{-}$) and can be used to quantify the edge flux $J_{\mathbb{i} \rightarrow \mathbb{j}}= \sum_{\mathcal{C}} J^+_{\mathcal{C}}-J^-_{\mathcal{C}}$, as the summation of the differences between the circuit fluxes along all cycle trajectories that include the edge $\mathbb{i} \rightarrow \mathbb{j}$. For example, the net edge flux $J_{\mathbb{2}\rightarrow\mathbb{3}}$ in Fig.~\ref{fig2} can be obtained by $J_{\mathbb{2}\rightarrow\mathbb{3}}=(J^+_{C_1}+J^+_{C_3}+J^+_{C_4})-(J^-_{C_1}+J^-_{C_3}+J^-_{C_4})$.

\begin{figure}[!h]
\centering
\includegraphics[width=\columnwidth]{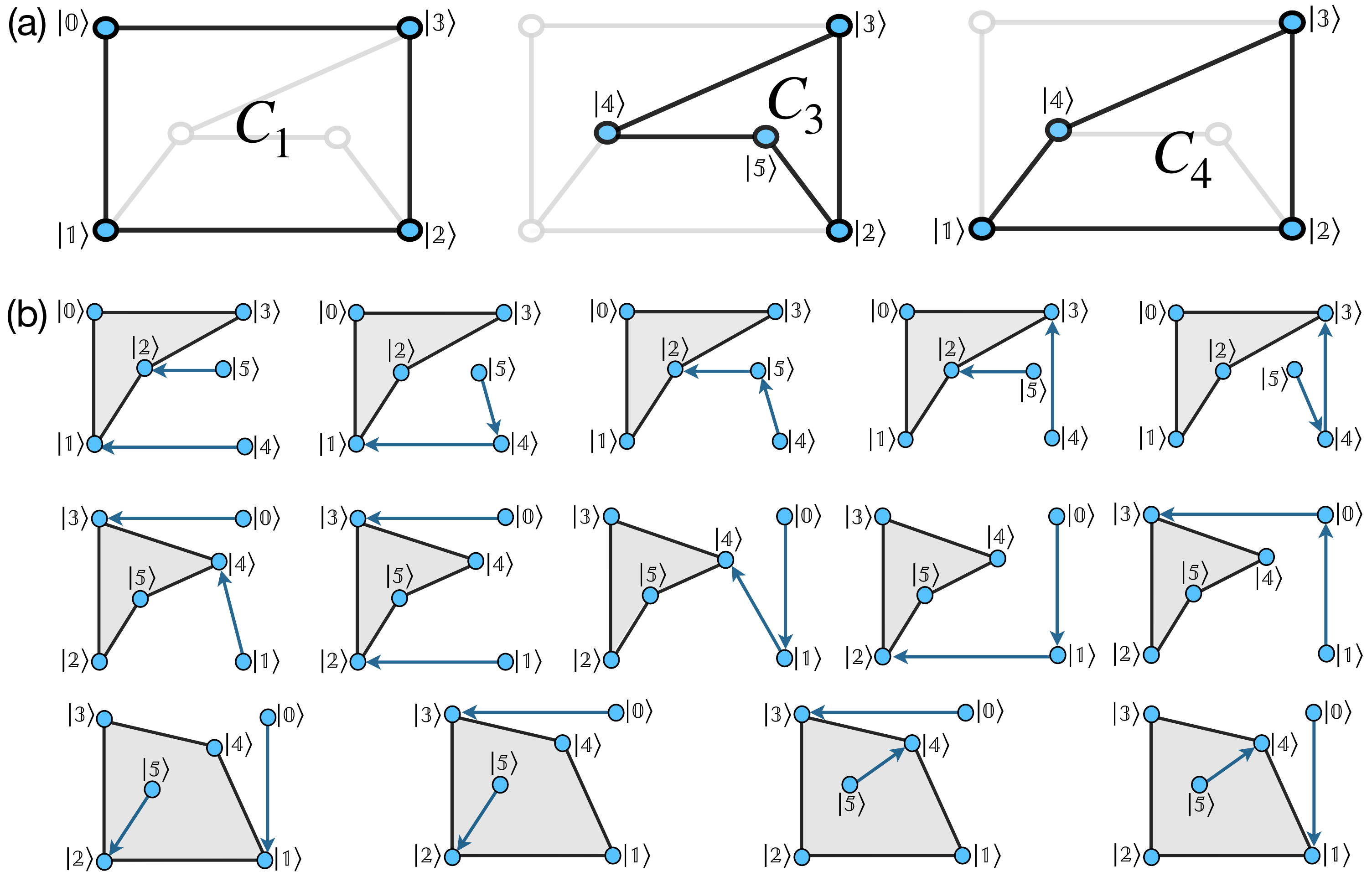}
\caption{~(a)~First three top-ranked cycle trajectories of our basic graph $\mathbb{G}$ shown in Fig.~\ref{fig2}. (b) Spanning trees rooted on the top-ranked cycles $C_1$, $C_3$, and $C_4$.}
\label{top_ranked_cycles_ST}
\end{figure}

Finally, the flux associated with each one-directional ``circuit'' is determined by~\cite{hill1975stochastic,kohler1980thefrequency,schnakenberg1976network} 
\begin{equation}
J^{\pm}_{C}=\Pi^{\pm}_{C}\frac{\Lambda_C}{\Lambda}.
\end{equation}
Here, $\Pi^{\pm}_{C}$ represents the weight factor which is defined by the product of transition rates along the cycle trajectory $C^{\pm}$, whereas the sum of the weights of the spanning trees rooted on cycle $C$ is given by $\Lambda_C$ and $\Lambda$ measures the total weight of the spanning trees rooted on each individual state. As an example, Fig~\ref{top_ranked_cycles_ST}, displays three such cycles of our basic graph $\mathbb{G}$ and the corresponding spanning trees that are rooted on them, where the cycle $C^+_1$ possesses a weight factor $\Pi^+_{C_1}=k_{\mathbb{03}}k_{\mathbb{32}}k_{\mathbb{21}}k_{\mathbb{10}}$. In reality, enumerating the vast number of spanning trees that are rooted in each individual vertices requires an inconceivable amount of work, particularly as the graph size escalates. Fortunately, we can navigate this problem by leveraging the generalized matrix-tree theorem from the algebraic graph~\cite{wang2022cycleflux,ren2017detectable}. Upon rewriting the master equation in the following form: $\frac{d\textbf{P}}{dt}=-\textbf{M}\textbf{P}$, where $\textbf{P}=\{P_\mathbb{0}, P_\mathbb{1}, P_\mathbb{2}, P_\mathbb{3}, P_\mathbb{4}, P_\mathbb{5}\}$ is a column matrix and 
\begin{widetext}
\begin{eqnarray}\label{laplacianMatrix}
\textbf{M}=\left[ {\begin{array}{cccccc}
   k_{\mathbb{10}}+k_{\mathbb{30}} & -k_{\mathbb{01}} & 0 & -k_{\mathbb{03}} & 0 & 0 \\
   -k_{\mathbb{10}} & k_{\mathbb{01}}+k_{\mathbb{41}}+k_{\mathbb{21}} & -k_{\mathbb{12}} & 0 & -k_{\mathbb{14}} & 0 \\
  0 & -k_{\mathbb{21}} & k_{\mathbb{12}}+k_{\mathbb{32}}+k_{\mathbb{52}} & -k_{\mathbb{23}} & 0 & -k_{\mathbb{25}} \\
   -k_{\mathbb{30}} & 0 & -k_{\mathbb{32}} & k_{\mathbb{03}}+k_{\mathbb{23}}+k_{\mathbb{43}} & -k_{\mathbb{34}} & 0 \\
   0 & -k_{\mathbb{41}} & 0 & -k_{\mathbb{43}} & k_{\mathbb{34}}+k_{\mathbb{14}}+k_{\mathbb{54}} & -k_{\mathbb{45}} \\
   0 & 0 & -k_{\mathbb{52}} & 0 & -k_{\mathbb{54}} & k_{\mathbb{25}}+k_{\mathbb{45}} \\
  \end{array} } \right],\nonumber\\
\end{eqnarray}
\end{widetext}
is the Laplacian (transition) matrix of the weighted graph, the matrix-tree theorem provides a powerful recipe for calculating the number of directed spanning trees rooted on a particular cycle.
Specifically, the matrix-tree theorem asserts that an effective expression for $\Lambda_C$ can be evaluated as the determinant of the principal minor of the Laplacian matrix $\textbf{M}$ of the basic graph, namely $\det(\textbf{M}[C; C])$. In other words, the determinant of the reduced matrix $\textbf{M}[C; C]$ obtained by removing the rows and columns indexed by $\mathbb{i}\in C$ of the original matrix $\textbf{M}$, is equal to the sum of the weights of directed spanning trees rooted on C, i.e., $\Lambda_C=\det(\textbf{M}[C; C])$. As an example, for the cycle $C_1$, the reduced Laplacian matrix $\textbf{M}[C_1;C_1]$ or $\textbf{M}[\mathbb{0,1,2,3;0,1,2,3}]$ and its determinant are given by:
\begin{eqnarray}
    &&\textbf{M}[C_1;C_1]=\left[ {\begin{array}{cc}
            k_{\mathbb{34}}+k_{\mathbb{14}}+k_{\mathbb{54}} & -k_{\mathbb{45}}  \\
           -k_{\mathbb{54}} & k_{\mathbb{25}}+k_{\mathbb{45}} \\
    \end{array} }\right], \\
    &&\det(\textbf{M}[C_1;C_1])= k_{\mathbb{25}}(k_{\mathbb{34}}+k_{\mathbb{14}}+k_{\mathbb{54}})+k_{\mathbb{45}}(k_{\mathbb{34}}+k_{\mathbb{14}}).\nonumber\\
\end{eqnarray}

Likewise, in the case of $\Lambda_\mathbb{i}$, the determinant related to the principal minor of the $\textbf{M}$ matrix can be derived by excluding the relevant row and column that correspond to state $\mathbb{i}$. Consequently, $\Lambda_\mathbb{i}$ can be represented by the determinant of $\textbf{M}[\mathbb{i}; \mathbb{i}]$ and the steady-state population, $\bar{P}_\mathbb{i}$, can be precisely written as the ratio of the determinants~\cite{wang2022cycleflux,ren2017detectable} 
\begin{equation}  
\bar{P}_\mathbb{i}=\frac{\Lambda_{\mathbb{i}}}{\Lambda} \equiv \frac{\det(\textbf{M}[\mathbb{i};\mathbb{i}])}{\sum_\mathbb{i} \det(\textbf{M}[\mathbb{i};\mathbb{i}])} ,
\end{equation}
where we identify $\Lambda$ as the sum of determinants of the principal minors of $\textbf{M}$, i.e., $\sum_\mathbb{i} \textbf{M}[\mathbb{i}; \mathbb{i}]$. Similarly, the one-directional circuit flux can directly be calculated as a product of two factors~\cite{wang2022cycleflux,ren2017detectable}
\begin{equation}\label{J_C_pm} 
J^{\pm}_{C}=\Pi^{\pm}_{C}\frac{\det(\textbf{M}[C;C])}{\sum_\mathbb{i} \det(\textbf{M}[\mathbb{i};\mathbb{i}])}. 
\end{equation}
The first coefficient captures the weight of the cycle in the particular direction, for instance, $C_{1} (\mathbb{0} \rightarrow \mathbb{1} \rightarrow \mathbb{2} \rightarrow \mathbb{3} \rightarrow \mathbb{0})$ in Fig~\ref{top_ranked_cycles_ST}(a), has a counterclockwise weight factor $\Pi^+_{C_1}=k_{\mathbb{03}}k_{\mathbb{32}}k_{\mathbb{21}}k_{\mathbb{10}}$. The subsequent term involves a ratio of two determinants: the numerator $\det(\textbf{M}[C_1 \{\mathbb{0},\mathbb{1},\mathbb{2},\mathbb{3}\}; C_1\{\mathbb{0},\mathbb{1},\mathbb{2},\mathbb{3}\}])$ tallies with the weighted summation of all five spanning trees rooted on $C_1 $ (Fig~\ref{top_ranked_cycles_ST}b), while the denominator $\sum_{\mathbb{i=0}}^{\mathbb{5}}\det(\textbf{M}[\mathbb{i};\mathbb{i}])$ serves as a constant of normalization factor via a common term, representing the total weight of the spanning trees rooted on each individual state. In summary, the graph-theoretic representation of the cycle flux can be intuitively understood as the flow of weighted edges on spanning trees directed towards a cycle, which is intricately linked to the frequencies at which the cycle trajectory occurs. As a result, the determinant $\det(\textbf{M}[C_1; C_1])$ exemplified above, can be alternatively computed by means of the determinant $\det(\textbf{N}[ \mathbb{k}; \mathbb{k}])$ of a ``new graph'' $\textbf{N}$ which is obtained by merging the set of vertices $\{\mathbb{0},\mathbb{1},\mathbb{2},\mathbb{3}\}$ into a new vertex ``$\mathbb{k}$'' within the initial graph $\textbf{M}$~\cite{wang2022cycleflux}, represented in Fig~\ref{top_ranked_cycles_ST}(b) by the shaded region.
Following the same procedure, we can evaluate the cycle fluxes of all the cycle trajectories of our basic graph and efficiently ranks out the top-ranked cycle fluxes. The generalized matrix-tree theorem thus provides valuable machinery by accomplishing the behavior of weighted graphs and their corresponding cycle fluxes. Notably, the proficiency of the cycle flux ranking scheme improves with the size of the graph~\cite{wang2022cycleflux}.

\section{Results and Discussion}\label{Sec-IV}
\begin{figure}[!h]
    \centering
    \includegraphics[width=\columnwidth]{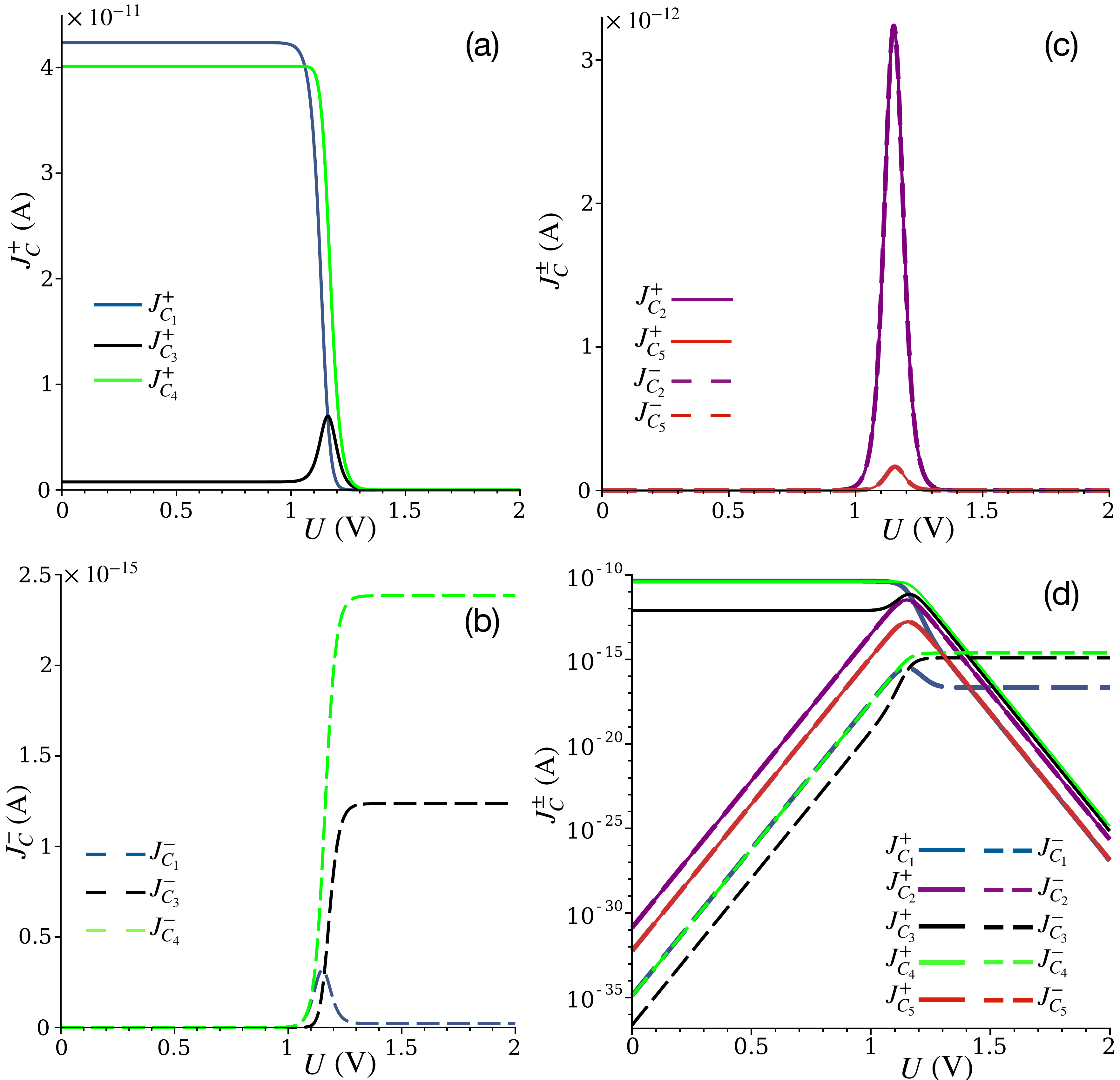}
    \caption{The circuit fluxes $J^\pm_C$ (in Ampere) as a function of bias voltage $U$ (in Volt) are plotted as follows: (a) top-ranked circuit fluxes $J^+_{C_1}$, $J^+_{C_3}$, $J^+_{C_4}$ vs. $U$ (b) lowest ranked circuit fluxes $J^-_{C_1}$, $J^-_{C_3}$, $J^-_{C_4}$ vs. $U$ (c) third and fourth-ranked paired circuits $J^\pm_{C_2}$ and $J^\pm_{C_5}$ vs. $U$, and (d) all circuit fluxes (in log scale) vs. $U$. The parameters are $\mu_{\rm L}=0$ ${\rm eV}$, $\mu_{\rm R}=\mu_{\rm L}+|e|U$, $\varepsilon_{\rm D1}=-0.1$ ${\rm eV}$, $\varepsilon_{\rm D2}=1.4$ ${\rm eV}$, $\varepsilon_{\rm A2}=0.9$ ${\rm eV}$, $U_{\rm A}=0.25$ ${\rm eV}$, $T_{\rm ph}=T_{\rm L}=T_{\rm R}=T$, $T=300$ K, $T_{\rm S}=6000$ K, $\gamma_{\rm L}=\gamma_{\rm R}=\gamma^{\rm D}_{\rm pht}=\gamma^{\rm D}_{\rm phn}=0.01\gamma^{\rm DA}_{\rm phn}$ and $\gamma^{\rm DA}_{\rm phn}=10^{12}$ $ {\rm s}^{-1}$~\cite{einax2011heterojunction}.}
    \label{top_ranked_cycle}
\end{figure}

In what follows, we utilize the aforementioned ranking scheme to examine the non-equilibrium transport characteristic of PV solar cells. To unfold the working principle of the donor-acceptor molecular junctions, we begin by decomposing the PV cell network into subcycles or a complete set of paired cycle trajectories. In the present case, the basic graph [Fig.~\ref{fig2}] consists of 5 subcycles or 10 paired one-directional circuits (cycle trajectories). Subsequently, the efficient cycle flux ranking scheme [Cf.~Eq.~\eqref{J_C_pm}] is implemented to rank out the top-ranked cycle fluxes and identify the major working cycles of the molecular solar cell. In Fig.~\ref{top_ranked_cycle}, we have plotted all the circuit fluxes against the bias voltage $U$ and notice that the cycle flux trajectories $C^+_1$ ($|\mathbb{1}\rangle\rightarrow|\mathbb{2}\rangle\rightarrow|\mathbb{3}\rangle\rightarrow|\mathbb{0}\rangle\rightarrow|\mathbb{1}\rangle$) or ($|1010\rangle\rightarrow|0110\rangle\rightarrow|0011\rangle\rightarrow|0010\rangle\rightarrow|1010\rangle$) and $C^+_4$ ($|\mathbb{1}\rangle\rightarrow|\mathbb{2}\rangle\rightarrow|\mathbb{3}\rangle\rightarrow|\mathbb{4}\rangle\rightarrow|\mathbb{1}\rangle$) or ($|1010\rangle\rightarrow|0110\rangle\rightarrow|0011\rangle\rightarrow|1011\rangle\rightarrow|1010\rangle$), are comparable in magnitudes and can be classified as the first-ranked circuits. The second-ranked circuit is found to be $C^+_3$ ($|\mathbb{3}\rangle\rightarrow|\mathbb{4}\rangle\rightarrow|\mathbb{5}\rangle\rightarrow|\mathbb{2}\rangle\rightarrow|\mathbb{3}\rangle$) or ($|0011\rangle\rightarrow|1011\rangle\rightarrow|0111\rangle\rightarrow|0110\rangle\rightarrow|0011\rangle$). Both the first and second-ranked circuits are plotted in Fig.~\ref{top_ranked_cycle}a, while $C^-_1$ ($|\mathbb{1}\rangle\rightarrow|\mathbb{0}\rangle\rightarrow|\mathbb{3}\rangle\rightarrow|\mathbb{2}\rangle\rightarrow|\mathbb{1}\rangle$) or ($|1010\rangle\rightarrow|0010\rangle\rightarrow|0011\rangle\rightarrow|0110\rangle\rightarrow|1010\rangle$), $C^-_3$ ($|\mathbb{3}\rangle\rightarrow|\mathbb{2}\rangle\rightarrow|\mathbb{5}\rangle\rightarrow|\mathbb{4}\rangle\rightarrow|\mathbb{3}\rangle$) or ($|0011\rangle\rightarrow|0110\rangle\rightarrow|0111\rangle\rightarrow|1011\rangle\rightarrow|0011\rangle$), and $C^-_4$ ($|\mathbb{1}\rangle\rightarrow|\mathbb{4}\rangle\rightarrow|\mathbb{3}\rangle\rightarrow|\mathbb{2}\rangle\rightarrow|\mathbb{1}\rangle$) or ($|1010\rangle\rightarrow|1011\rangle\rightarrow|0011\rangle\rightarrow|0110\rangle\rightarrow|1010\rangle$), are the lowest ranked cycle trajectories shown in Fig.~\ref{top_ranked_cycle}b. On the contrary, the third-ranked paired circuits $C^{\pm}_2$ correspond to sequence of states ($|\mathbb{1}\rangle\leftrightarrow|\mathbb{4}\rangle\leftrightarrow|\mathbb{3}\rangle\leftrightarrow|\mathbb{0}\rangle\leftrightarrow|\mathbb{1}\rangle$) or ($|1010\rangle\leftrightarrow|1011\rangle\leftrightarrow|0011\rangle\leftrightarrow|0010\rangle\leftrightarrow|1010\rangle$), and the fourth-ranked paired circuits, $C^{\pm}_5$ correspond to sequence of states ($|\mathbb{1}\rangle\leftrightarrow|\mathbb{2}\rangle\leftrightarrow|\mathbb{5}\rangle\leftrightarrow|\mathbb{4}\rangle\leftrightarrow|\mathbb{1}\rangle$) or ($|1010\rangle\leftrightarrow|0110\rangle\leftrightarrow|0111\rangle\leftrightarrow|1011\rangle\leftrightarrow|1010\rangle$), are plotted in Fig.~\ref{top_ranked_cycle}c. From Fig.~\ref{top_ranked_cycle}d, we can understand that circuit fluxes corresponding to $C^{\pm}_2$ and $C^{\pm}_5$ are not only several orders of magnitude smaller than the first ($C^{+}_1$ and $C^{+}_4$) and second-ranked ($C^{+}_3$) cycle trajectories, they also have an equal amount of flux current in both directions. Therefore, the cycle affinity~\cite{einax2016maximum}, $\mathcal{A}$ becomes zero upon the addition of two counter-pair cycle trajectories, where the cycle affinity is defined as $\mathcal{A}=-\ln{\mathcal{K}}$, $\mathcal{K}={\Pi^{+}_{C}}/{\Pi^{-}_{C}}$ being the ratio of the forward and the backward rates for a specific cycle. Since the magnitude of the cycle flux is the difference between the circuit fluxes $J_C=J^{+}_{C}-J^{-}_{C}$, it follows immediately that cycles $C_1$, $C_3$ and $C_4$ possess nonzero flux current fulfilling $\mathcal{K}\neq 1$, while cycle flux associated with $C_2$ and $C_5$ are identically zero, satisfying~$\mathcal{K}=1$.

\begin{figure}[!h]
\centering
\includegraphics[width=1\columnwidth]{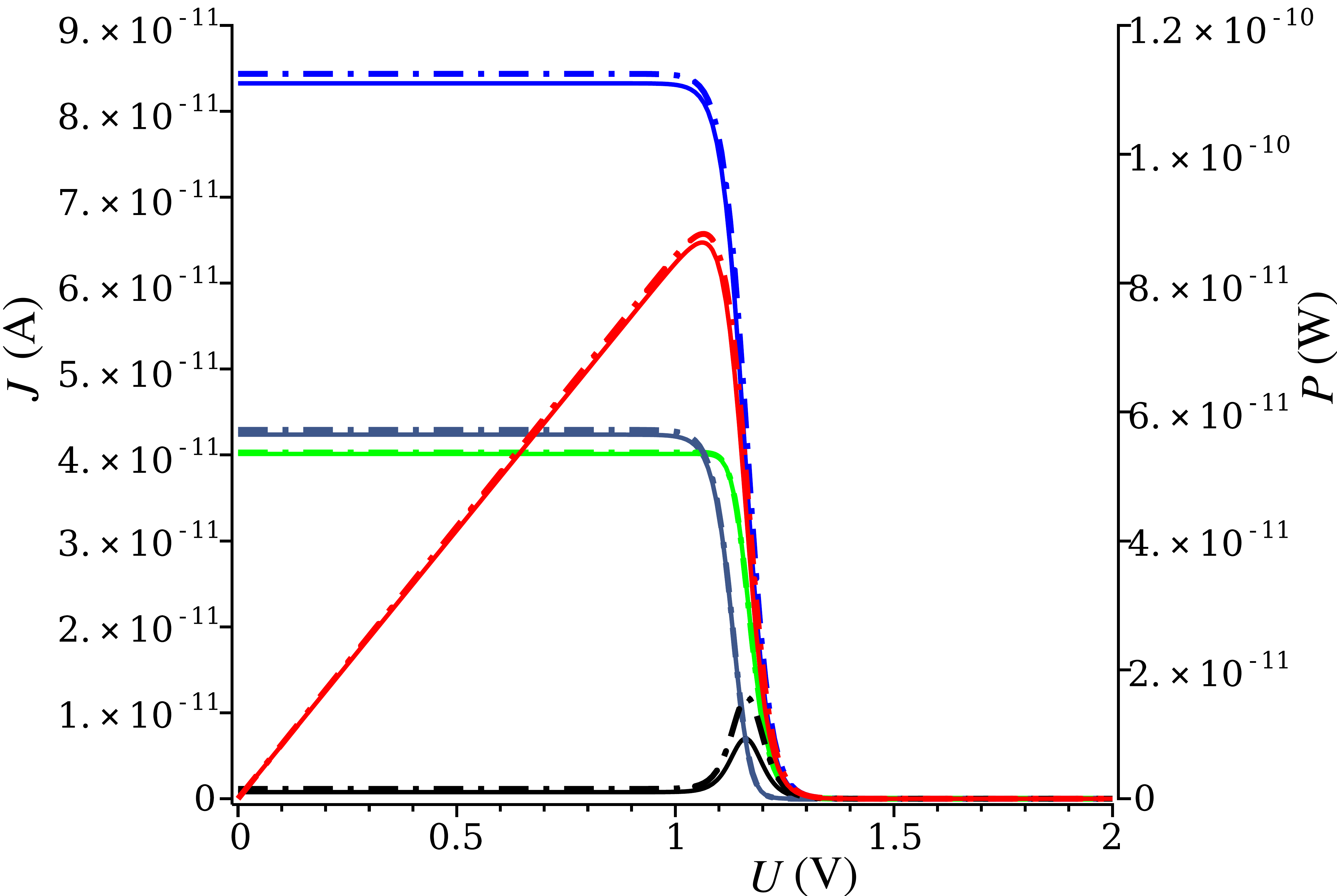}
\caption{Dual axis plot of top-ranked cycle fluxes $J_{C_1}$ (azure blue), $J_{C_3}$ (black), $J_{C_4}$ (green), and electron current $J$ (blue) on the left vertical axis (in Ampere) as a function of voltage bias $U$ and $P$ (red) on the right vertical axis (in Watt). The parameters are the same as mentioned in Fig.~\ref{top_ranked_cycle}. The solid lines represents $J(P)$ in the presence of nonradiative loss process ($\gamma^{\rm D}_{\rm phn}=10^{10}$ $ {\rm s}^{-1}$) whereas the dash-dot lines represents $J(P)$ in the absence of nonradiative losses ($\gamma^{\rm D}_{\rm phn}=0$).}
\label{J(P)_vs_U}
\end{figure}

As a result, we conclude that the $C_2$ and $C_5$ cycles do not contribute to the overall photocurrent, and the entire contribution to the electron current generated within the photocell solely comes from the first ($C_1$ and $C_4$) and second-ranked ($C_3$) cycles. We have made a dual-axis plot to illustrate the total current ($J(U)$) and the corresponding power ($P=UJ(U)$), where $P$ passes through a maximum in Fig.~\ref{J(P)_vs_U}, as expected. Our analysis reveals that the total current is really the sum of the top three cycle fluxes, i.e, $J=J_{C_1}+J_{C_4}+J_{C_3}$, regardless of the presence or the absence of the nonradiative loss processes due to $\gamma^{\rm D}_{\rm phn}$. To make a proper comparison, we have taken the parameter set used by Nitzan's group~\cite{einax2011heterojunction} and found excellent agreement with their results. Qualitatively, one can understand that among the five possible cycles, those cycles ($C_1$, $C_3$ and $C_4$) which involve the $|\mathbb{2} \rangle \leftrightarrow |\mathbb{3} \rangle$ transition, contribute a finite amount to the overall electron current, whereas sub-cycles $C_2$ and $C_5$ which do not possess this edge, have zero contribution to the total current. This can be attributed to the fact that among various possible electron transfer channels, those pathways are only relevant that do involve the electron transfer between donor and acceptor molecules. In graph-theoretical language, this corresponds to subcycles with nonzero edge flux along $|\mathbb{2}\rangle \leftrightarrow |\mathbb{3}\rangle$. Furthermore, it is evident from Fig.~\ref{J(P)_vs_U} that compared to first-ranked cycles, the second-ranked cycle $C_3$ makes a negligible contribution to the overall electron current for the majority of the parameter range. Consequently, the total flux current $J$ is simply twice that of the two individual first-ranked cycles. Intuitively, one expects the $C_1$ electron transfer channel to be the one and only exclusive route for the electron transfer pathways in photovoltaic devices~\cite{einax2014network,einax2016maximum}. Surprisingly, our analysis reveals the existence of an equally important $C_4$ pathway that contributes to a similar extent to the overall photocurrent, along with the natural $C_1$ electron transfer channel. These findings represent the first set of important results for our efficient ranking scheme, which goes beyond the conventional understanding of the electron transfer pathways in molecular photovoltaics.

\begin{figure}
    \centering
    \includegraphics[width=0.95\columnwidth]{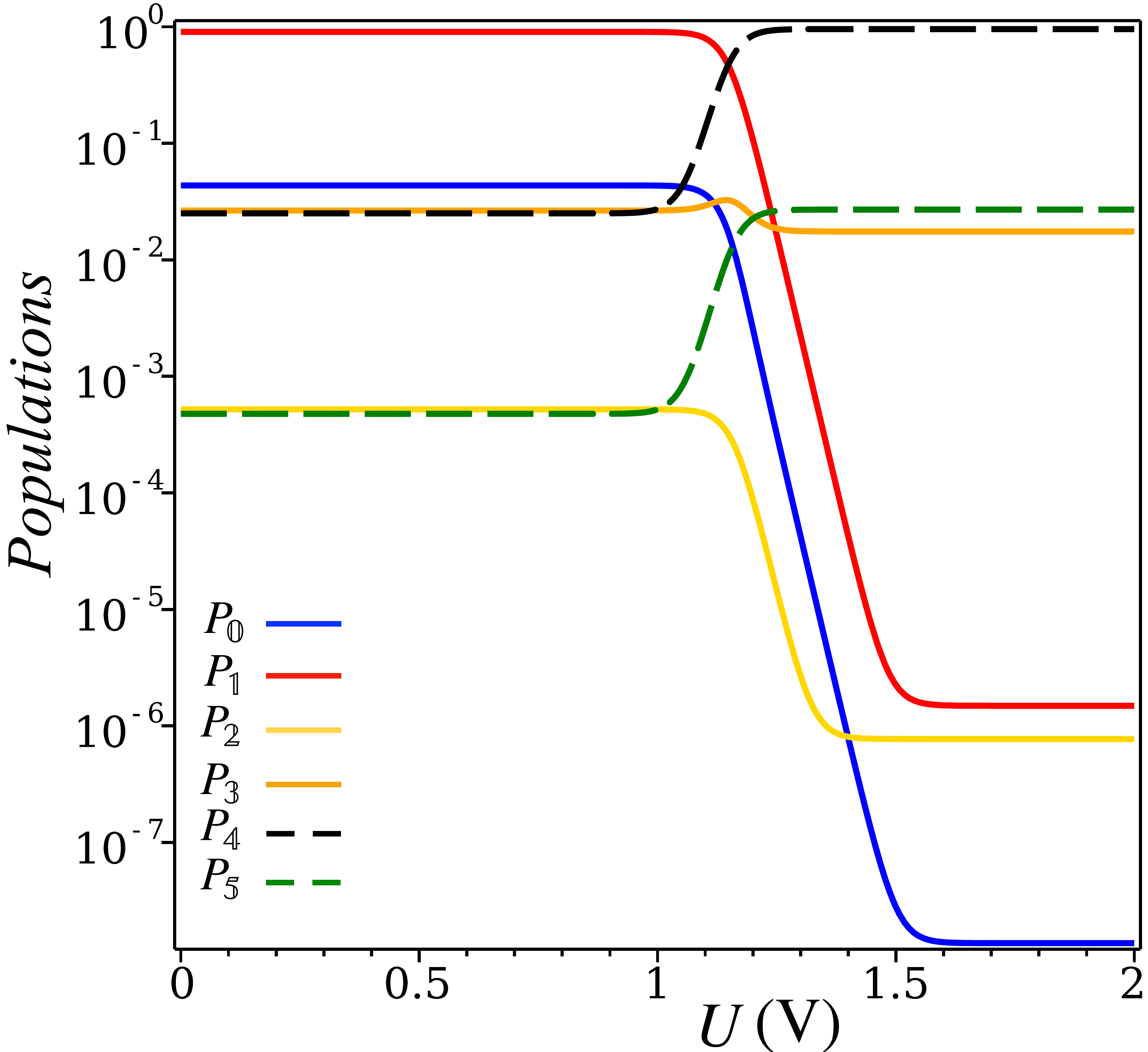}
    \caption{Populations of the different states as a function of $U$ (in Volts). The parameters are $\mu_{\rm L}=0$ ${\rm eV}$, $\mu_{\rm R}=\mu_{\rm L}+|e|U$, $\varepsilon_{\rm D1}=-0.1$ ${\rm eV}$, $\varepsilon_{\rm D2}=1.4$ ${\rm eV}$, $\varepsilon_{\rm A2}=0.9$ ${\rm eV}$, $U_{\rm A}=0.25$ ${\rm eV}$, $T_{\rm ph}=T_{\rm L}=T_{\rm R}=T$, $T=300$ K, $T_{\rm S}=6000$ K, $\gamma_{\rm L}=\gamma_{\rm R}=\gamma^{\rm D}_{\rm pht}=\gamma^{\rm D}_{\rm phn}=0.01\gamma^{\rm DA}_{\rm phn}$ and $\gamma^{\rm DA}_{\rm phn}=10^{12}$ $ {\rm s}^{-1}$~\cite{einax2011heterojunction}.}
    \label{populations}
\end{figure}

Secondly, it can be inferred from Fig.\ref{J(P)_vs_U} that $C_1$ and $C_4$ operate as first-ranked cycles, while $C_3$ acts as a second-ranked cycle for bias voltage $U$ around $\sim 1$ eV. After that, both the net current as well as the power experience a simultaneous drop. The explanation for this behavior can be acquired from the population plot shown in Fig.\ref{populations}, as a function of $U$. For bias voltages $U \sim 1$ eV, $|1010\rangle$ or $|\mathbb{1}\rangle$ is the maximally populated state, and both the highest-ranked cycles start with the initial state $|1010\rangle$, represented by the curve $P_{\mathbb{1}}$ in Fig.~\ref{populations}. In contrast, the second-ranked cycle $C_3$ starts with a less populated initial state $|0011\rangle$, denoted by $P_{\mathbb{0}}$. Notably, the first two steps of the highest-ranked cycles are identical: it starts with the electron transfer between the levels ${\rm D1}$ to ${\rm D2}$ at the donor site that involve radiative (photon-induced) and nonradiative (phonon-induced) processes, while the second step designates the electron transfer between donor and acceptor molecules. It is worth pointing out that the above two steps of the highest-ranked cycles represent their counter-clockwise (forward) cycle trajectories, namely, $C^+_1$ and $C^+_4$, respectively. This is reasonable since the backward (clockwise) cycle flux trajectories are negligible compared to their forward counterpart and because of this reason it is the counter-clockwise or forward cycle trajectories that predominately determines the underlying directions of the cycle flux of the individual subcycles. 
In particular, the dynamical steps of the cycle $C^+_1$ are as follows: starting from the neutral state $|1010\rangle$ ($|\mathbb{1}\rangle$),  the system transits sequentially into $|0110\rangle$ ($|\mathbb{2}\rangle$) by absorbing the photon (electron transfer between ${\rm D1}$ and ${\rm D2}$ at the donor site), followed by $|\mathbb{3}\rangle$ or $|0011\rangle$ (via phonon relaxation), $|\mathbb{0}\rangle$ or $|0010\rangle$ (one electron tunnels from ${\rm A2}$ of the acceptor into the right electron reservoir) and finally returns to its initial state $|\mathbb{1}\rangle$ or $|1010\rangle$ (where one electron tunnels from the left reservoir into ${\rm D1}$ of the donor). Similarly, for cycle, $C^+_4$, the first two processes are common with the cycle $C^+_1$ (i.e., $|1010\rangle\rightarrow|0110\rangle$ ($|\mathbb{1}\rangle\rightarrow|\mathbb{2}\rangle$) and $|0110\rangle\rightarrow|0011\rangle$ ($|\mathbb{2}\rangle\rightarrow|\mathbb{3}\rangle$)). The system then changes its state from $|0011\rangle$ to $|1011\rangle$ or $|\mathbb{3}\rangle\rightarrow|\mathbb{4}\rangle$, i.e., one electron tunnel from the left reservoir into ${\rm D1}$ of the donor, and finally, returns backs to its initial state $|\mathbb{1}\rangle$ or $|1010\rangle$ (where one electron tunnels from the ${\rm A2}$ of the acceptor into the right reservoir). So, these are the two pertinent electron transfer pathways that play equivalent roles in a typical donor-acceptor molecular junction photocell. For all the numerical plots presented in Figs.~\ref{top_ranked_cycle}-~\ref{populations}, we have used the same set the parameters as Nitzan et. al.~\cite{einax2011heterojunction}, which are also reasonable from the experimental perspective as reported in Ref.~\cite{rice1996theory,soci2007photoconductivity,pensack2010beyond}.

To understand the significant drop in the total electron current and power, after a certain bias voltage in Fig.~\ref{J(P)_vs_U}, we need to examine the steady-state populations as shown in Fig.~\ref{populations}. We observe that as long as the chemical potential of the right reservoir, $\mu_{\rm R}$, reaches the energy level of $\varepsilon_{\rm A2}$, the trend of the populations obeys $P_\mathbb{1}>>P_\mathbb{0}>P_\mathbb{3}\simeq P_\mathbb{4}>P_\mathbb{2}\simeq P_\mathbb{5}$. However, once $\mu_{\rm R}$ exceeds the energy of the acceptor level ${\rm A2}$, the probability of finding the system in states $|1010\rangle$ ($P_{\mathbb{1}}$), $|0010\rangle$ ($P_{\mathbb{0}}$), and $|0110\rangle$ ($P_{\mathbb{2}}$) decreases sharply. Meanwhile, the probability of finding the system in states $|1011\rangle$ ($P_{\mathbb{4}}$) and $|0111\rangle$ ($P_{\mathbb{3}}$) increases rapidly. This indicates that electron tunneling from the excited state of the acceptor into the right reservoir becomes less favorable. As a result, the circuit fluxes $J^+_{C_1}$ and $J^+_{C_4}$ remain constant until $\mu_{\rm R}$ reaches the energy level of $\varepsilon_{\rm A2}$, and then decrease steadily after surpassing this energy level. We can also analyze the second-ranked circuit flux, $J^+_{C_3}$, which remains constant until $\mu_{\rm R}$ reaches the energy level $\varepsilon_{\rm A2}$, and becomes maximum precisely at the midpoint of $\varepsilon_{\rm D2}$ and $\varepsilon_{\rm A2}$, i.e., $1.15$ ${\rm eV}$. By carefully analyzing Fig.~\ref{J(P)_vs_U}, we can see that we have varied the bias voltages by changing $\mu_{\rm R}$ while keeping all other parameters fixed. Now, Fig.\ref{fig2} shows that $\mu_{\rm R}$ governs transitions between pairs of states where the acceptor's excited state is occupied in at least one of them. For the counter-clock $C_1$ cycle, the second last edge ($|\mathbb{3}\rangle \leftrightarrow |\mathbb{0}\rangle$) is controlled by $\mu_{\rm R}$, whereas, for $C_4$, it is the last edge ($|\mathbb{4}\rangle \leftrightarrow |\mathbb{1}\rangle$) that is controlled by the bias voltage. Thus, we get a single drop in the current vs. voltage diagram as $\mu_{\rm R}$ crosses the value of $\varepsilon_{\rm A2}$. In the case of the $C_3$ cycle, the last two edges (($|\mathbb{4}\rangle \leftrightarrow |\mathbb{5}\rangle$) and ($|\mathbb{5}\rangle \leftrightarrow |\mathbb{2}\rangle$)) are controlled by $\mu_{\rm R}$. Since $\mu_{\rm R}$ controls both transitions, the cycle current will be maximum when both transitions are favored optimally~\cite{shuvadip2022univarsal}, i.e., at the average value of both energy levels, $\varepsilon_{\rm D2}$ and $\varepsilon_{\rm A2}$. For the present choice of parameters, this turns out to be precisely $1.15$ eV, as obtained in Fig.~\ref{J(P)_vs_U}.

\begin{figure}
    \centering
    \includegraphics[width=0.9\columnwidth]{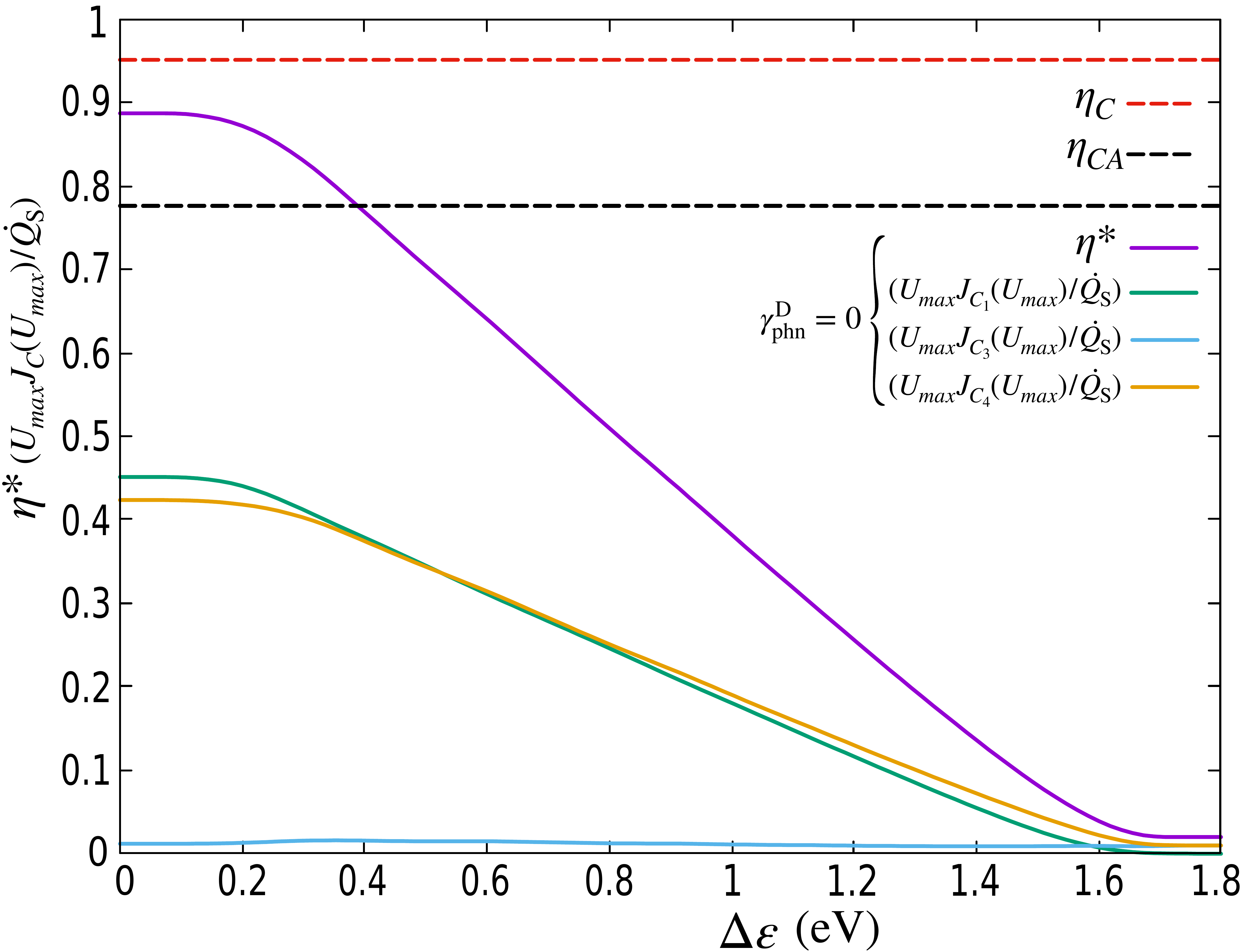}
    \caption{Plot of $\eta^*$ ($U_{max}J_{C}(U_{max})/\dot{Q}_{\rm S}$) vs $\Delta\varepsilon$ (eV) in absence of any nonradiative loss processes and the parameters are $\mu_{\rm L}=0$ ${\rm eV}$, $\mu_{\rm R}=\mu_{\rm L}+|e|U$, $\varepsilon_{\rm D1}=-0.1$ ${\rm eV}$, $\varepsilon_{\rm D2}=1.4$ ${\rm eV}$, $\varepsilon_{\rm A2}=\varepsilon_{\rm D2}-\Delta\varepsilon$, $U_{\rm A}=0.25$ ${\rm eV}$, $T_{\rm ph}=T_{\rm L}=T_{\rm R}=T$, $T=300$ K, $T_{\rm S}=6000$ K, $\gamma_{\rm L}=\gamma_{\rm R}=\gamma^{\rm D}_{\rm pht}=0.01\gamma^{\rm DA}_{\rm phn}$, and $\gamma^{\rm DA}_{\rm phn}=10^{12}$ $ {\rm s}^{-1}$~\cite{einax2011heterojunction}. Red and black dashed lines represent the Carnot and Curzon-Ahlborn bounds respectively.}
    \label{efficiency_plot2}
\end{figure}
\begin{figure}
    \centering
    \includegraphics[width=0.9\columnwidth]{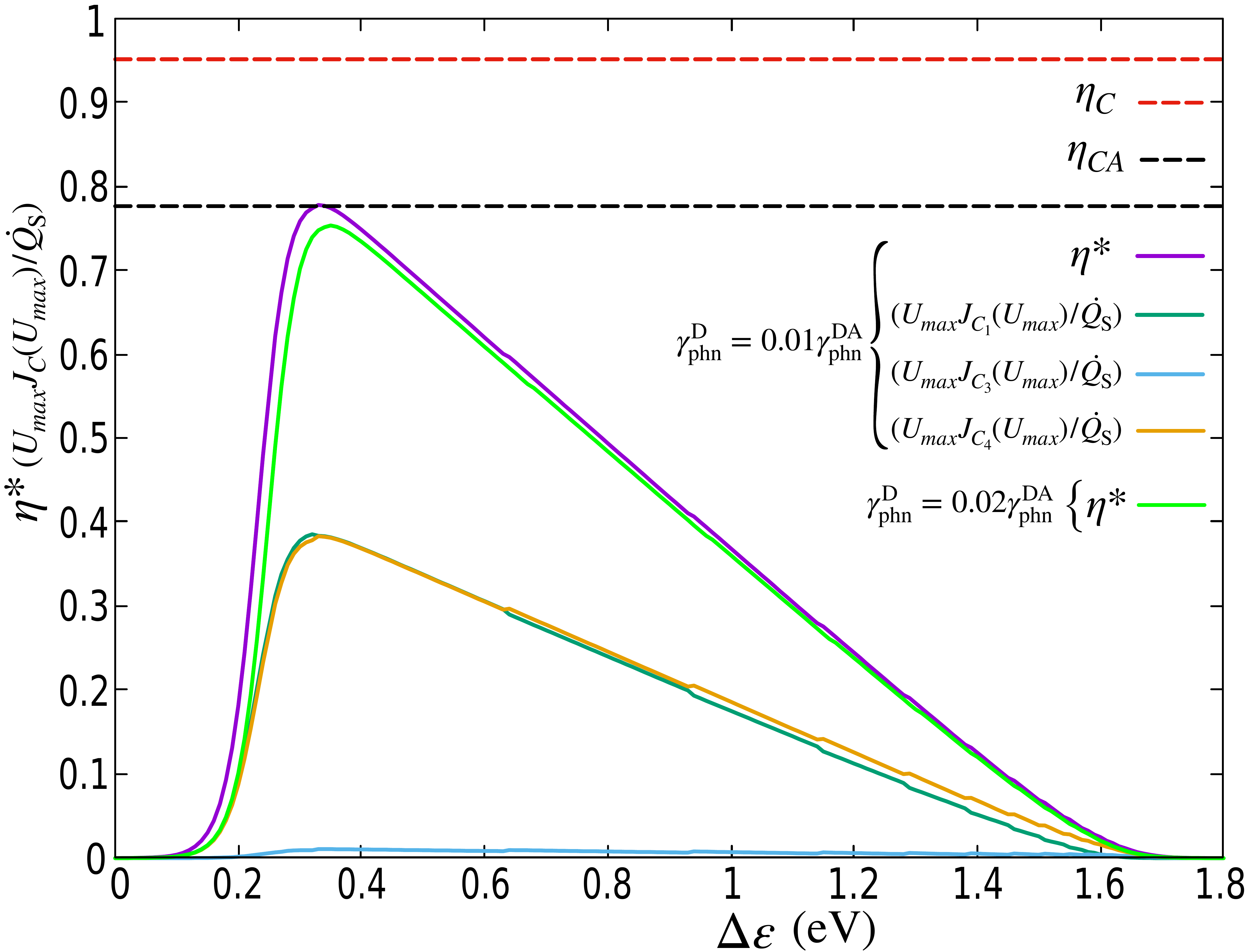}
    \caption{Plot of $\eta^*$ ($U_{max}J_{C}(U_{max})/\dot{Q}_{\rm S}$) vs $\Delta\varepsilon$ (eV) in presence of nonradtive losses arising from $\gamma^{\rm D}_{\rm {phn}}$ and the parameters are $\mu_{\rm L}=0$ ${\rm eV}$, $\mu_{\rm R}=\mu_{\rm L}+|e|U$, $\varepsilon_{\rm D1}=-0.1$ ${\rm eV}$, $\varepsilon_{\rm D2}=1.4$ ${\rm eV}$, $\varepsilon_{\rm A2}=\varepsilon_{\rm D2}-\Delta\varepsilon$, $U_{\rm A}=0.25$ ${\rm eV}$, $T_{\rm ph}=T_{\rm L}=T_{\rm R}=T$, $T=300$ K, $T_{\rm S}=6000$ K, $\gamma_{\rm L}=\gamma_{\rm R}=\gamma^{\rm D}_{\rm pht}=0.01\gamma^{\rm DA}_{\rm phn}$ and $\gamma^{\rm DA}_{\rm phn}=10^{12}$ $ {\rm s}^{-1}$~\cite{einax2011heterojunction}. Red and black dashed lines represent the Carnot and Curzon-Ahlborn bounds respectively.}
    \label{efficiency_plot1}
\end{figure}
From the above analysis, it becomes clear that the donor-acceptor energy gap plays a crucial role in determining the performance of the photocell. To this end, in Figs.~\ref{efficiency_plot2} and~\ref{efficiency_plot1}, we consider for our photocell, a real practical quantity of interest, i.e.,  the efficiency at the maximum power as a function of the donor-acceptor energy gap $\Delta \varepsilon=\varepsilon_{\rm D2}-\varepsilon_{\rm A2}$. To be precise, we define the thermodynamic efficiency~\cite{einax2011heterojunction,rutten2009reaching,giebink2011thermodynamic} at maximum power as follows:
\begin{eqnarray}
\label{efficiency_def}
\eta^* &=&\frac{P_{max}}{\dot{Q}_{\rm S}}=\frac{U_{max}J(U_{max})}{\dot{Q}_{\rm S}},\\
  &\approx& \frac{U_{max}J_{C_1}(U_{max})}{\dot{Q}_{\rm S}} +\frac{U_{max}J_{C_4}(U_{max})}{\dot{Q}_{\rm S}}\label{eff_def2},
\end{eqnarray}
where $U_{max}$ denotes the bias voltage at which power passes through a maximum, i.e., at $U=U_{max}$, power becomes $P_{max}=U_{max}J(U_{max})$. In Eq.~\eqref{efficiency_def}, the denominator $\dot{Q}_{\rm S}=\Delta E J_{\rm S}(U_{max})$, represents the total energy absorbed per unit time from the radiation field, with $\Delta E$ being the energy gap between the donor excited and ground levels. We rewrite the expression of efficiency at maximum power in terms of the two first-ranked cycle fluxes in Eq.~\eqref{eff_def2}, where each term represents the contribution of its respective cycle to the overall efficiency. From Figs.~\ref{efficiency_plot2} and~\ref{efficiency_plot1}, we can conclude that the cycles $C_1$ and $C_4$ are the primary electron transfer channels across the entire parameter space, regardless of whether radiative process or non-radiative losses are at play. Interestingly, the plots exhibit a distinctive peak in the behavior of $\eta^*$, in the presence of non-radiative loss processes. In such situations, $\eta^*$ attains a maximum value as a function of $\Delta \varepsilon$, where the energy gap $\Delta\varepsilon$ regulates the electron transfer process between donor and acceptor [Fig.~\ref{efficiency_plot1}]. Moreover, it is worth noting that although $\eta^*$ is bounded by the Carnot ($\eta_C=1-T/T_{\rm S}$), it can exceed the standard Curzon-Ahlborn bound~\cite{curzon1975efficiency} ($\eta_{CA}=1-\sqrt{T/T_{\rm S}}$) in the absence of any nonradiative losses [Fig.~\ref{efficiency_plot2}]. In other words, the strength of nonradiative loss processes is found to have a deleterious effect on the overall performance of the solar cell. It may even reduce the efficiency at maximum power at a value well below the $\eta_{CA}$ as shown in Fig.~\ref{efficiency_plot1}. Hence, in order to improve the device performance to its ultimate level, it becomes crucial to diligently mitigate all forms of nonradiative losses to a considerable extent.

\section{Conclusion}\label{Sec.V}

We introduce the top-ranked cycle flux ranking scheme of network analysis as a tool to elucidate the complex working principles of molecular junction solar cells. Our approach takes advantage of the mapping between the dissipative Lindblad master equation for molecular systems and the quantum-transition network that characterizes the nonequilibrium transport behavior of molecular photocells. We now summarize our key insights as follows:

(i) We have provided a microscopic Hamiltonian description of the phenomenological rate equations that are commonly used to characterize molecular junction solar cells. Based on a minimal model Hamiltonian, \textit{classical} looking rate equations are derived from detailed quantum Lindblad master equations. The resulting rate equations, though appear ``classical", the underlying transition rates are shown to be quantum mechanical in nature.

(ii) With the implementation of the effective ranking scheme, we have predicted the existence of a counterintuitive electron transfer pathway, which provides valuable insights into the detailed working principles of molecular photocells. Through rigorous analysis, we have clearly justified that cycles that contribute to the overall photocurrent must involve nonzero edge flux between donor and acceptor molecules embedded in a complex graph of quantum transition networks. We provide explanations for the drop in the current and power when measured against the bias voltage, as well as the various cycle fluxes. Our result advances the conventional understanding of nonequilibrium electron transfer pathways in donor-acceptor molecular junction solar cells which might be a significant step toward making efficient photovoltaic devices in the near future.

(iii) Finally, we obtain a crucial insight into the efficiency of photocells, revealing that their maximum power efficiency in the absence of non-radiative losses, can exceed the conventional Curzon-Ahlborn bound, yet abide by the Carnot limit. Nonetheless, the incorporation of a non-radiative recombination process at the donor site has been found to have a detrimental effect on the photocell's performance, reducing its efficiency at maximum power, below the Curzon-Ahlborn limit. These findings underscore the intricacies of various factors that govern the overall performance of molecular photovoltaics. While in our present work, top-rankled cycle fluxes capture the essential physics of molecular junction solar cells, future research can take into account other additional effects, such as environment-assisted electron transfer and recombination rates~\cite{subhajit2020environment}, Marcus homogeneous and heterogeneous electron transfer rates~\cite{einax2013multiple}, from a similar perspective. Thus the current approach offers important insights for further research in the field of photovoltaics. 

\section*{Acknowledgement}
AG acknowledges financial support from the Initiation grant of IITK (Grant No. IITK/CHM/2018513). N.G. is thankful to CSIR for the fellowship, and it is supported in part by the International Centre for Theoretical Sciences (ICTS) during a visit for participating in the online program  -  Bangalore School on Statistical Physics XII  (code: ICTS/bssp2021/6). S.G. is grateful to the Ministry of Education, Government of India, for the Prime Minister Research Fellowship (PMRF).

\newpage
\onecolumngrid 
\appendix

\section{Derivation of the Lindblad Master Equation and the electron currents}\label{Appendix-A}

The total Hamiltonian of the system interacting with the environments (baths) can be written as
\begin{equation}
    H_T=H_{el}+H_B+H_{SB},
\end{equation}
where $H_{el}$ is the Hamiltonian of the molecular system and $H_B$ is the total bath Hamiltonian which is given by $H_B= H_{\rm L}+H_{\rm R}+H_{\rm pht}+H_{\rm phn}$. The interaction Hamiltonian between different baths and molecular sites is given by
\begin{equation}
    H_{SB}=H_I+H^{\rm pht}_I+H^{\rm phn}_I.
\end{equation}
Here, $H_I=H^{\rm L}_I+H^{\rm R}_I$ represents the interaction of the molecular sites (donor HOMO and acceptor LUMO) with the left and right electrodes. $H^{\rm pht}_I$ characterizes the interaction of the photon bath with the donor site only, and $H^{\rm phn}_I$ denotes the interaction of the phonon bath with the respective molecular sites~\cite{ajisaka2015themolecular,subhajit2020environment,agarwalla2015fullcounting}. For convenience, we rewrite the interaction Hamiltonians [Eqs.~\eqref{H_I}-\eqref{phonon_H_int}] of the main text as
\begin{eqnarray}
    H^{\rm L}_I=\sum_l \hbar g_l(d^\dagger_l c_{\rm D1}+c^\dagger_{\rm D1} d_l); \quad
    H^{\rm R}_I=\sum_r \hbar g_r(d^\dagger_r c_{\rm A2}+c^\dagger_{\rm A2} d_r), \quad
    H^{\rm pht}_I = \sum_k \hbar g^{\rm D}_{k}(a^\dagger_k c^\dagger_{\rm D1}c_{\rm D2}+c^\dagger_{\rm D2}c_{\rm D1} a_k),\nonumber\\
    H^{\rm phn}_I = \sum_q \hbar g^{\rm D}_{q}(b^\dagger_q c^\dagger_{\rm D1}c_{\rm D2}+c^\dagger_{\rm D2}c_{\rm D1} b_q)
    + \sum_q\hbar g^{\rm DA}_{q}(b^\dagger_q c^\dagger_{\rm A2}c_{\rm D2}+c^\dagger_{\rm D2}c_{\rm A2} b_q). 
\end{eqnarray}
In order to derive the master equation, one can start with the von Neumann equation for the total density matrix $\rho_T$ in the interaction picture which is given by 
\begin{equation}
    \frac{d\rho_T}{dt}=-\frac{i}{\hbar}[H_{SB}(t),\rho_T(t)].
\end{equation}
Under the Born-Markov approximation, the master equation in terms of the reduced density matrix $\rho$ of the system can be written as~\cite{breuer2002book,gupt2022PRE,shuvadip2022univarsal} 
\begin{equation}\label{masterEq}
    \frac{d\rho(t)}{dt}=-\frac{1}{\hbar^2}\Tr_B\int^\infty_0 ds [H_{SB}(t),[H_{SB}(t-s),\rho_T(t)]],
\end{equation}
where $\rho(t)=\Tr_{B}\{\rho_T\}\equiv\Tr_{{\rm L},{\rm R},{\rm pht},{\rm phn}}\{\rho(t)\otimes\rho_{\rm L}\otimes\rho_{\rm R}\otimes\rho_{\rm pht}\otimes\rho_{\rm phn}\}$ and $\Tr_{{\rm L},{\rm R},{\rm pht},{\rm phn}}$ stands for the trace over each bath degrees of freedom. As a result, we can rewrite the Eq.~\eqref{masterEq} as
\begin{equation}\label{ME_rho}
   \frac{d\rho(t)}{dt}=-\frac{1}{\hbar^2}\Tr_{{\rm L},{\rm R},{\rm pht},{\rm phn}}\int^\infty_0 ds [H_{SB}(t),[H_{SB}(t-s),\rho(t)\otimes\rho_{\rm L}\otimes\rho_{\rm R}\otimes\rho_{\rm pht}\otimes\rho_{\rm phn}]]. 
\end{equation}
Since the bath operators obey the following relations
\begin{eqnarray}
   \Tr_{\rm L}(d_l(t) \rho_{\rm L})&=&0= \Tr_{\rm L}(d^\dagger_l(t) \rho_{\rm L}), \quad\quad\quad \Tr_{\rm R}(d_r(t) \rho_{\rm R})=0= \Tr_{\rm R}(d^\dagger_r(t) \rho_{\rm R}),\nonumber\\
   \Tr_{\rm pht}(a_k(t) \rho_{\rm pht})&=&0= \Tr_{\rm pht}(a^\dagger_k(t) \rho_{\rm pht}), \quad \Tr_{\rm phn}(b_q(t) \rho_{\rm phn})=0= \Tr_{\rm phn}(b^\dagger_q(t) \rho_{\rm phn}),
\end{eqnarray}
one can verify that~\cite{gupt2022PRE,shuvadip2022univarsal}
\begin{equation}
    \Tr_{{\rm L},{\rm R},{\rm pht},{\rm phn}}\{ [H^\alpha_I(t),[H^\beta_{I}(t-s),\rho(t)\otimes\rho_{\rm L}\otimes\rho_{\rm R}\otimes\rho_{\rm pht}\otimes\rho_{\rm phn}]]\}=0;\quad \alpha\ne\beta, \quad \alpha, \beta={\rm L},{\rm R},{\rm pht},{\rm phn}.
\end{equation}
This simplifies Eq.~\eqref{ME_rho} to 
\begin{equation}
    \frac{d\rho(t)}{dt}=-\frac{1}{\hbar^2}\sum_\alpha\Tr_{{\rm L},{\rm R},{\rm pht},{\rm phn}}\int^\infty_0 ds [H^\alpha_{I}(t),[H^\alpha_{I}(t-s),\rho(t)\otimes\rho_{\rm L}\otimes\rho_{\rm R}\otimes\rho_{\rm pht}\otimes\rho_{\rm phn}]].
\end{equation}
In the above equation, we use the interaction picture system operators as~\cite{breuer2002book}
\begin{eqnarray}
    c_{\rm si}(t)&=&e^{{iH_{el} t}/{\hbar}}c_{\rm si} e^{{-iH_{el} t}/{\hbar}}=\sum_{\{\varepsilon_{\mathbb{ji}}\}} e^{{-i{\varepsilon_{\mathbb{ji}}} t}/{\hbar}}c_{\rm si}; \nonumber\\
    c^\dagger_{\rm si}(t)&=&e^{{iH_{el} t}/{\hbar}}c^\dagger_{\rm si} e^{{-iH_{el} t}/{\hbar}}=\sum_{\{\varepsilon_{\mathbb{ji}}\}} e^{{i{\varepsilon_{\mathbb{ji}}} t}/{\hbar}}c^\dagger_{\rm si}; \quad\quad{\rm s =D, A} \quad \text{and}\quad {\rm i}=1,2;
\end{eqnarray}
where $\varepsilon_{\mathbb{ji}}=\varepsilon_{\mathbb{j}}-\varepsilon_{\mathbb{i}}>0$ is the transition energy associated with the $\mathbb{j}$-th and  $\mathbb{i}$-th states of the molecular system. Similarly, one can also evaluate the expressions for the interaction picture bath operators~\cite{breuer2002book}. With the above prescriptions, we can derive after a little bit of algebra, the compact form of the following Lindblad master equation 
\begin{equation}\label{Lindblad_MEqn}
  \frac{d\rho}{dt}=\mathcal{L}_{\rm L}[\rho]+\mathcal{L}_{\rm R}[\rho]+\mathcal{L}^{\rm D}_{\rm pht}[\rho] + \mathcal{L}^{\rm D}_{\rm phn}[\rho]+\mathcal{L}^{\rm DA}_{\rm phn}[\rho].  
\end{equation}
The explicit forms of the Lindblad superoperator $\mathcal{L}$ in the above equation are given by
\begin{eqnarray}\label{Lindblad_L}
   \mathcal{L}_{\rm L}[\rho]=\sum_{\{\varepsilon_{\rm L}\}} \gamma_{\rm L}\Big[ f(\varepsilon_{\rm L},\mu_{\rm L},T_{\rm L})\Big(c^\dagger_{\rm D1}(\varepsilon_{\rm L})\rho c_{\rm D1}(\varepsilon_{\rm L})-\frac{1}{2}\{c_{\rm D1}(\varepsilon_{\rm L})c^\dagger_{\rm D1}(\varepsilon_{\rm L}),\rho\}\Big) \nonumber\\
   + (1-f(\varepsilon_{\rm L},\mu_{\rm L},T_{\rm L}))\Big(c_{\rm D1}(\varepsilon_{\rm L})\rho c^\dagger_{\rm D1}(\varepsilon_{\rm L})-\frac{1}{2}\{c^\dagger_{\rm D1}(\varepsilon_{\rm L})c_{\rm D1}(\varepsilon_{\rm L}),\rho\}\Big) \Big],
\end{eqnarray}
\begin{eqnarray}\label{Lindblad_R}
   \mathcal{L}_{\rm R}[\rho]=\sum_{\{\varepsilon_{\rm R}\}} \gamma_{\rm R}\Big[ f(\varepsilon_{\rm R},\mu_{\rm R},T_{\rm R})\Big(c^\dagger_{\rm A2}(\varepsilon_{\rm R})\rho c_{\rm A2}(\varepsilon_{\rm R})-\frac{1}{2}\{c_{\rm A2}(\varepsilon_{\rm R})c^\dagger_{\rm A2}(\varepsilon_{\rm R}),\rho\}\Big) \nonumber\\
   + (1-f(\varepsilon_{\rm R},\mu_{\rm R},T_{\rm R}))\Big(c_{\rm A2}(\varepsilon_{\rm R})\rho c^\dagger_{\rm A2}(\varepsilon_{\rm R})-\frac{1}{2}\{c^\dagger_{\rm A2}(\varepsilon_{\rm R})c_{\rm A2}(\varepsilon_{\rm R}),\rho\}\Big) \Big],
\end{eqnarray}
\begin{eqnarray}\label{Lindblad_pht}
   \mathcal{L}^{\rm D}_{\rm pht}[\rho]=\sum_{\{\varepsilon_{\rm D}\}} \gamma^{\rm D}_{\rm pht}\Big[ n(\varepsilon_{\rm D},T_{\rm S})\Big(V^\dagger_{\rm D}(\varepsilon_{\rm D})\rho V_{\rm D}(\varepsilon_{\rm D})-\frac{1}{2}\{V_{\rm D}(\varepsilon_{\rm D})V^\dagger_{\rm D}(\varepsilon_{\rm D}),\rho\}\Big) \nonumber\\
   + (n(\varepsilon_{\rm D},T_{\rm S})+1)\Big(V_{\rm D}(\varepsilon_{\rm D})\rho V^\dagger_{\rm D}(\varepsilon_{\rm D})-\frac{1}{2}\{V^\dagger_{\rm D}(\varepsilon_{\rm D})V_{\rm D}(\varepsilon_{\rm D}),\rho\}\Big) \Big],
\end{eqnarray}
\begin{eqnarray}\label{Lindblad_phn_D}
   \mathcal{L}^{\rm D}_{\rm phn}[\rho]=\sum_{\{\varepsilon_{\rm D}\}} \gamma^{\rm D}_{\rm phn}\Big[ n(\varepsilon_{\rm D},T_{\rm ph})\Big(V^\dagger_{\rm D}(\varepsilon_{\rm D})\rho V_{\rm D}(\varepsilon_{\rm D})-\frac{1}{2}\{V_{\rm D}(\varepsilon_{\rm D})V^\dagger_{\rm D}(\varepsilon_{\rm D}),\rho\}\Big) \nonumber \\
   + (n(\varepsilon_{\rm D},T_{\rm ph})+1)\Big(V_{\rm D}(\varepsilon_{\rm D})\rho V^\dagger_{\rm D}(\varepsilon_{\rm D})-\frac{1}{2}\{V^\dagger_{\rm D}(\varepsilon_{\rm D})V_{\rm D}(\varepsilon_{\rm D}),\rho\}\Big) \Big],
\end{eqnarray}
\begin{eqnarray}\label{Lindblad_phn_DA}
   \mathcal{L}^{\rm DA}_{\rm phn}[\rho]=\sum_{\{\varepsilon_{\rm DA}\}} \gamma^{\rm DA}_{\rm phn}\Big[ n(\varepsilon_{\rm DA},T_{\rm ph})\Big(V^\dagger_{\rm DA}(\varepsilon_{\rm DA})\rho V_{\rm DA}(\varepsilon_{\rm DA})-\frac{1}{2}\{V_{\rm DA}(\varepsilon_{\rm DA})V^\dagger_{\rm DA}(\varepsilon_{\rm DA}),\rho\}\Big) \nonumber\\
   + (n(\varepsilon_{\rm DA},T_{\rm ph})+1)\Big(V_{\rm DA}(\varepsilon_{\rm DA})\rho V^\dagger_{\rm DA}(\varepsilon_{\rm DA})-\frac{1}{2}\{V^\dagger_{\rm DA}(\varepsilon_{\rm DA})V_{\rm DA}(\varepsilon_{\rm DA}),\rho\}\Big) \Big],
\end{eqnarray}
where we define the operators, $\{V^{\dagger}_{\rm D}=c^\dagger_{\rm D2} c_{\rm D1}$, $V_{\rm D}=c^\dagger_{\rm D1} c_{\rm D2}\}$ and $\{V^{\dagger}_{\rm DA}=c^\dagger_{\rm D2} c_{\rm A2}$, $V_{\rm DA}=c^\dagger_{\rm A2} c_{\rm D2}\}$, as the combination of system operators that are responsible for the transition between donor ground and excited states, as well as the transition between the excited states of the donor and acceptor molecules, respectively. The electron transfer rates corresponding to their respective bath are characterized by the various $\gamma$'s. Their explicit forms in terms of the system-reservoir coupling constants can be calculated by Fermi's golden rule, as $\gamma_{L(R)}=\sum_{l(r)}2\pi\hbar|g_{l(r)}|^2 \delta\big(\varepsilon-\epsilon_{l(r)}\big)$, $\gamma^D_{\rm pht}=\sum_k 2\pi\hbar |g^D_k|^2 \delta\big(\varepsilon-\epsilon_{k}\big)$, and $\gamma^r_{\rm phn}=\sum_q 2\pi\hbar |g^r_q|^2 \delta\big(\varepsilon-\epsilon'_{q}\big)$ where $r={\rm D}$, ${\rm DA}$. The function $f(\varepsilon,\mu, T)=[e^{(\varepsilon-\mu)/k_B T}+1]^{-1}$ is the Fermi-Dirac distribution for the left ($\rm L$) and right ($\rm R$) bath with energy $\varepsilon$, chemical potential $\mu$ and temperature $T$ respectively. Analogously, $n(\varepsilon, T)=[e^{\varepsilon/k_B T}-1]^{-1}$ is the Bose-Einstein distribution corresponding to the photon and phonon baths with energy $\varepsilon$ and temperature $T$ respectively. The distribution functions are obtained by tracing over the respective bath density operator. For examples, $\Tr_{\rm L(R)}\big(d^\dagger_{l(r)} d_{l(r)} \rho_{\rm L(R)}\big)=f\big(\epsilon_{l(r)},\mu_{\rm L(R)}, T_{\rm L(R)}\big)$, and $\Tr_{\rm L(R)}\big(d_{l(r)} d^\dagger_{l(r)} \rho_{\rm L(R)}\big)=1-f\big(\epsilon_{l(r)},\mu_{\rm L(R)}, T_{\rm L(R)}\big)$, where the bath operators $d^\dagger_{l(r)}$ and $d_{l(r)}$ obey anti-commutation relation, whereas, $\Tr_{\rm pht}\big(a^\dagger_{k} a_{k} \rho_{\rm pht}\big)=n\big(\epsilon_{k}, T_{\rm S}\big)$, $\Tr_{\rm pht}\big(a_{k} a^\dagger_{k} \rho_{\rm pht}\big)=1+n\big(\epsilon_{k}, T_{\rm S}\big)$, and $\Tr_{\rm phn}\big(b^\dagger_{q} b_{q} \rho_{\rm phn}\big)=n\big(\epsilon'_{q}, T_{\rm ph}\big)$, $\Tr_{\rm phn}\big(b_{q} b^\dagger_{q} \rho_{\rm phn}\big)=1+n\big(\epsilon'_{q}, T_{\rm ph}\big)$ where the operators $a^\dagger_k$ ($b^\dagger_q$) and $a_k$ ($b_q$) follow commutation relations. In both cases, $k_B$ is the Boltzmann constant and the transition energies  driven by the $\rm L (R)$ baths are $\varepsilon_{\rm L}=\varepsilon_{\mathbb{10}},\varepsilon_{\mathbb{43}}$, and $\varepsilon_{\rm R}=\varepsilon_{\mathbb{30}},\varepsilon_{\mathbb{41}},\varepsilon_{\mathbb{52}}$ respectively, while  $\varepsilon_{\rm D}=\varepsilon_{\mathbb{21}},\varepsilon_{\mathbb{54}}$ (driven by both photon and phonon bath), and $\varepsilon_{\rm DA}=\varepsilon_{\mathbb{23}}$ (driven by phonon bath). Finally, we note that the creation and annihilation operators can be expressed in terms of the system eigenstates in the following form
\begin{eqnarray}\label{bra_ket_notation}
    c^\dagger_{\rm D1}&=&|\mathbb{1}\rangle\langle \mathbb{0}|+|\mathbb{4}\rangle\langle \mathbb{3}|; \quad c_{\rm D1}=|\mathbb{0}\rangle\langle \mathbb{1}|+|\mathbb{3}\rangle\langle \mathbb{4}|, \nonumber\\
    c^\dagger_{\rm D2}&=&|\mathbb{2}\rangle\langle \mathbb{0}|+|\mathbb{5}\rangle\langle \mathbb{3}|; \quad c_{\rm D2}=|\mathbb{0}\rangle\langle \mathbb{2}|+|\mathbb{3}\rangle\langle \mathbb{5}|, \nonumber\\
    c^\dagger_{\rm A2}&=&|\mathbb{3}\rangle\langle \mathbb{0}|+|\mathbb{4}\rangle\langle \mathbb{1}|+|\mathbb{5}\rangle\langle \mathbb{2}|, \nonumber\\ c_{\rm A2}&=&|\mathbb{0}\rangle\langle \mathbb{3}|+|\mathbb{1}\rangle\langle \mathbb{4}|+|\mathbb{2}\rangle\langle \mathbb{5}|.
\end{eqnarray}
So, the time evolution of the occupation probabilities which are the diagonal elements of the reduced density matrix $P_\mathbb{i}=\langle \mathbb{i}|\rho|\mathbb{i} \rangle$, can be obtained using the Lindblad master equation [Eq.~\eqref{Lindblad_MEqn}] in the following way. For example, 
\begin{eqnarray}\label{EOM_P0}
    \frac{dP_\mathbb{0}}{dt}=\langle \mathbb{0}|\frac{d\rho}{dt}|\mathbb{0}\rangle =\langle \mathbb{0}|\mathcal{L}_{\rm L}[\rho]|\mathbb{0}\rangle+\langle \mathbb{0}|\mathcal{L}_{\rm R}[\rho]|\mathbb{0}\rangle+\langle \mathbb{0}|\mathcal{L}^{\rm D}_{\rm pht}[\rho]|\mathbb{0}\rangle + \langle \mathbb{0}|\mathcal{L}^{\rm D}_{\rm phn}[\rho]|\mathbb{0}\rangle 
    +\langle \mathbb{0}|\mathcal{L}^{\rm DA}_{\rm phn}[\rho]|\mathbb{0}\rangle,
\end{eqnarray}
and the first term of the R.H.S. can be calculated by using Eqs.~\eqref{Lindblad_L} and \eqref{bra_ket_notation} as:
\begin{eqnarray}
    \langle \mathbb{0}|\mathcal{L}_{\rm L}[\rho]|\mathbb{0}\rangle&&= \langle \mathbb{0}|\sum_{\{\varepsilon_{\rm L}\}} \gamma_{\rm L}\Big[ f(\varepsilon_{\rm L},\mu_{\rm L},T_{\rm L})\Big(c^\dagger_{\rm D1}\rho c_{\rm D1}-\frac{1}{2}\{c_{\rm D1}c^\dagger_{\rm D1},\rho\}\Big) 
   + (1-f(\varepsilon_{\rm L},\mu_{\rm L},T_{\rm L}))\Big(c_{\rm D1}\rho c^\dagger_{\rm D1}-\frac{1}{2}\{c^\dagger_{\rm D1}c_{\rm D1},\rho\}\Big) \Big]|\mathbb{0}\rangle \nonumber\\
   &&={\gamma_{\rm L} f(\varepsilon_{\mathbb{10}},\mu_{\rm L},T_{\rm L})}\Big(\langle \mathbb{0}|\mathbb{1}\rangle\langle \mathbb{0}|\rho|\mathbb{0}\rangle\langle \mathbb{1}|\mathbb{0}\rangle -\frac{1}{2}\langle \mathbb{0}|\mathbb{0}\rangle\langle \mathbb{0}|\rho|\mathbb{0}\rangle -\frac{1}{2} \langle \mathbb{0}|\rho|\mathbb{0}\rangle\langle \mathbb{0}|\mathbb{0}\rangle\Big) \nonumber\\
   &&+ \gamma_{\rm L} (1-f(\varepsilon_{\mathbb{10}},\mu_{\rm L},T_{\rm L})) \Big(\langle \mathbb{0}|\mathbb{0}\rangle\langle \mathbb{1}|\rho|\mathbb{1}\rangle\langle \mathbb{0}|\mathbb{0}\rangle -\frac{1}{2}\langle \mathbb{0}|\mathbb{1}\rangle\langle \mathbb{1}|\rho|\mathbb{0}\rangle -\frac{1}{2} \langle \mathbb{0}|\rho|\mathbb{1}\rangle\langle \mathbb{1}|\mathbb{0}\rangle \Big) \nonumber\\
   &&=\gamma_{\rm L} (1-f(\varepsilon_{\mathbb{10}},\mu_{\rm L},T_{\rm L})) P_\mathbb{1} - \gamma_L f(\varepsilon_{\mathbb{10}},\mu_{\rm L},T_{\rm L}) P_\mathbb{0} 
   \equiv  k_{\mathbb{01}}P_\mathbb{1}-k_{\mathbb{10}}P_\mathbb{0}.
\end{eqnarray}
Here we identify $\varepsilon_{\mathbb{10}}=\varepsilon_\mathbb{1} -\varepsilon_\mathbb{0}=\varepsilon_{\rm D1}$. Similarly, one can check that the second term reduces to
\begin{eqnarray}
   \langle \mathbb{0}|\mathcal{L}_{\rm R}[\rho]|\mathbb{0}\rangle &&= \gamma_{\rm R} (1-f(\varepsilon_{\mathbb{30}},\mu_{\rm R},T_{\rm R}))\langle \mathbb{3}|\rho|\mathbb{3}\rangle - \gamma_{\rm R} f(\varepsilon_{\mathbb{30}},\mu_{\rm R},T_{\rm R}) \langle \mathbb{0}|\rho|\mathbb{0}\rangle \nonumber\\
   &&= \gamma_{\rm R} (1-f(\varepsilon_{\mathbb{30}},\mu_{\rm R},T_{\rm R})) P_\mathbb{3} - \gamma_{\rm R} f(\varepsilon_{\mathbb{30}},\mu_{\rm R},T_{\rm R}) P_\mathbb{0} 
   \equiv k_{\mathbb{03}} P_\mathbb{3} - k_{\mathbb{30}} P_\mathbb{0},
\end{eqnarray}
and last three terms $\langle \mathbb{0}|\mathcal{L}^{\rm D}_{\rm pht}[\rho]|\mathbb{0}\rangle$, $\langle \mathbb{0}|\mathcal{L}^{\rm D}_{\rm phn}[\rho]|\mathbb{0}\rangle$ and $\langle \mathbb{0}|\mathcal{L}^{\rm DA}_{\rm phn}[\rho]|\mathbb{0}\rangle$ are equal to zero. Combing these results, we obtain from Eq.~\eqref{EOM_P0}
\begin{equation}
   \frac{dP_\mathbb{0}}{dt}=(k_{\mathbb{01}}P_\mathbb{1}-k_{\mathbb{10}}P_\mathbb{0})+(k_{\mathbb{03}}P_\mathbb{3}-k_{\mathbb{30}}P_\mathbb{0}), 
\end{equation}
which corresponds to our main text Eq.~\eqref{eq_P0} and the corresponding transition rates are summarized in Eqs.~\eqref{rate_kL}-\eqref{rate_kDA} of the main text. Following a similar procedure, we can derive all the main text Eqs.~\eqref{eq_P1}-\eqref{eq_P5} for the population dynamics. In terms of the occupation probabilities of the individual states, phenomenological expressions for the electron currents can be written following Nitzan et.al~\cite{einax2011heterojunction} as 
\begin{eqnarray}
    J_{\rm L} &=& [k_{\mathbb{10}}P_0-k_{\mathbb{01}}P_\mathbb{1}]+[k_{\mathbb{43}}P_\mathbb{3}-k_{\mathbb{34}}P_\mathbb{4}],\\
    J_{\rm R} &=& [k_{\mathbb{30}}P_\mathbb{0}-k_{\mathbb{03}}P_\mathbb{3}]+[k_{\mathbb{41}}P_\mathbb{1}-k_{\mathbb{14}}P_\mathbb{4}]+[k_{\mathbb{52}}P_\mathbb{2}-k_{\mathbb{25}}P_\mathbb{5}],\\
    J_{\rm S} &=& k_{\rm r}[P_\mathbb{1}+P_\mathbb{4}]-\Tilde{k}_{\rm r}[P_\mathbb{2}+P_\mathbb{5}], \\
    J_{\rm nr}&=&  k_{\rm nr}[P_\mathbb{1}+P_\mathbb{4}]-\Tilde{k}_{\rm nr}[P_\mathbb{2}+P_\mathbb{5}],  \\
    J_{\rm DA}&=& k_{23}P_\mathbb{3} - k_{\mathbb{\mathbb{32}}}P_\mathbb{2}.
\end{eqnarray}
Here $J_{\rm L}(J_{\rm R})$ is the electron currents entering (leaving) the molecular system from (to) the electrodes, $J_{\rm S}$ and $J_{\rm nr}$ are, respectively, the radiative (photon-induced) and nonradiative (phonon induced losses) electron currents between ground and excited states of the donor, and $J_{\rm DA}$ is the average current due to transfer of electrons between donor and acceptor species. Following Ref.~\cite{subhajit2020environment}, one can derive the same from the microscopic picture, starting from the Lindblad master equation Eq.~\eqref{Lindblad_MEqn}. In particular, steady-state electron currents can be obtained by using the following definitions  
\begin{eqnarray}
    J_{\rm L}&=&\Tr{N_{\rm L} \mathcal{L}_{\rm L}[\rho]},\\
    J_{\rm R}&=&\Tr{N_{\rm R} \mathcal{L}_{\rm R}[\rho]},\\
    J_{\rm S}&=&\Tr{N_{\rm D} \mathcal{L}^{\rm D}_{\rm pht}[\rho]},\\
    J_{\rm nr}&=&\Tr{N_{\rm D} \mathcal{L}^{\rm D}_{\rm phn}[\rho]},\\
    J_{\rm DA}&=&\Tr{N_{\rm DA} \mathcal{L}^{\rm DA}_{\rm phn}[\rho]}.
\end{eqnarray}
of current where $N_{\rm K}$ (${\rm K=L}$, ${\rm R}$, ${\rm D}$, ${\rm DA}$) is the number operator which is defined as
\begin{eqnarray}\label{number_operators}
    N_{\rm L}&=& c^\dagger_{\rm D1}c_{\rm D1}=|\mathbb{1}\rangle\langle \mathbb{1}|+|\mathbb{4}\rangle\langle \mathbb{4}|,\nonumber\\
    N_{\rm R}&=& c^\dagger_{\rm A2}c_{\rm A2}=|\mathbb{3}\rangle\langle \mathbb{3}|+|\mathbb{4}\rangle\langle \mathbb{4}|+|\mathbb{5}\rangle\langle \mathbb{5}|,\nonumber\\
    N_{\rm D}&=& V^\dagger_{\rm D}V_{\rm D}= c^\dagger_{\rm D2}c_{\rm D1}c^\dagger_{\rm D1}c_{\rm D2}=|\mathbb{2}\rangle\langle \mathbb{2}|+|\mathbb{5}\rangle\langle \mathbb{5}|,\nonumber\\
    N_{\rm DA}&=& V^\dagger_{\rm DA}V_{\rm DA}= c^\dagger_{\rm D2}c_{\rm A2}c^\dagger_{\rm A2}c_{\rm D2}=|\mathbb{2}\rangle\langle \mathbb{2}|.
\end{eqnarray}

\end{document}